\documentclass[preprint2]{emulateapj}
\usepackage{natbib}
\usepackage{hhline, amsmath}
\usepackage{graphicx}
\usepackage{xspace, times}
\usepackage{hyperref}
\usepackage{color, comment}

\newcommand{\um}{\,$\mu$m\xspace}

\newcommand{\nustar}{\emph{NuSTAR}\xspace}

\newcommand{\sbat}{\emph{Swift}/BAT\xspace}

\newcommand{\wise}{\emph{WISE}\xspace}
\newcommand{\herschel}{\emph{Herschel}\xspace}

\shorttitle{IR and X-ray Analysis of \sbat AGN}
\shortauthors{Lanz et al.}

\begin{document}

\title{\vspace{-6mm}Investigating the Covering Fraction Distribution of \sbat AGNs \\with X-ray and IR observations}

\author{Lauranne Lanz$^{1}$, Ryan\,C. Hickox$^{1}$, Mislav Balokovi\'{c}$^{2}$, 
Taro Shimizu$^{3}$, Claudio Ricci$^{4,5,6}$, Andy D. Goulding$^{7}$,  \\ David\,R. Ballantyne$^{8}$, Franz\,E. Bauer$^{4,9,10}$, Chien-Ting\,J. Chen$^{11}$, Agnese del Moro$^{3}$, Duncan Farrah$^{12}$, Michael, J. Koss$^{13}$, Stephanie LaMassa$^{14}$, Alberto Masini$^{1,15,16}$,  Luca Zappacosta$^{17}$ }

\affil{
$^{1}${Department of Physics and Astronomy, Dartmouth College, 6127 Wilder Laboratory, Hanover, NH 03755, USA}\\
$^{2}${Harvard-Smithsonian Center for Astrophysics, 60 Garden Street, Cambridge, MA 02138, USA} \\
$^{3}${Max-Planck Institute for Extra-terrestrial Physics, Giessenbachstrasse 1, D-85741 Garshing, Germany}\\
$^{4}${N\'ucleo de Astronom\'ia de la Falcultad de Ingenier\'ia, Unversidad Diego Potales, Av. Ej\'ercito Libertador 441, Santiago, Chile}\\
$^{5}${Kavli Institute for Astronomy and Astrophysics, Peking University, Beijing 100871, People's Republic of China}\\
$^{6}${Chinese Academy of Sciences South America Center for Astronomy, Camino El Observatorio 1515, Las Condes, Santiago, Chile}\\
$^{7}${Department of Astrophysics, Princeton University, Princeton, NJ 08544,  USA}\\
$^{8}${Center for Relativistic Astrophysics, School of Physics, Georgia Institute of Technology, 837 State Street, Atlanta, GA 30332, USA}\\
$^{9}${Millennium Institute of Astrophysics (MAS), Nuncio Monse{\~{n}}or S{\'{o}}tero Sanz 100, Providencia, Santiago, Chile} \\
$^{10}${Space Science Institute, 4750 Walnut Street, Suite 205, Boulder, Colorado 80301, USA} \\
$^{11}${Department of Astronomy and Astrophysics, Pennsylvania State University, 514 Davey Lab, University Park, PA 16802, USA}\\
$^{12}${Department of Physics, Virginia Tech, 850 West Campus Drive, Blacksburg, VA 24061, USA} \\
$^{13}${Eureka Scientific Inc., 2542 Delmar Avenue, Suite 100, Oakland, CA, 94602-3017, USA}\\
$^{14}${Space Telescope Science Institute, 3700 San Martin Dr., Baltimore, MD 21218, USA}\\
$^{15}${INAF-Osservatorio Astronomico di Bologna, via Gobetti 93/3, I-40129 Bologna, Italy}\\
$^{16}${Dipartimento di Fisica e Astronomia (DIFA), Universita di Bologna, via Gobetti 93/2, I-40129 Bologna, Italy}\\
$^{17}${INAF - Osservatorio Astronomico di Roma, via di Frascati 33, I-00078 Monte Porzio Catone, Italy}
}

\begin{abstract}
We present an analysis  of a sample of 69 local obscured {\em Swift}/Burst Alert Telescope active galactic nuclei (AGNs) with X-ray spectra from {\em NuSTAR}  and infrared (IR) spectral energy distributions from {\em{Herschel}} and {\em{WISE}}. We combine this X-ray and IR phenomenological modeling and find a significant correlation between reflected hard X-ray emission and IR AGN emission,  with suggestive indications that this correlation may be stronger than the one between intrinsic hard X-ray and IR emissions.  This relation between the IR and reflected X-ray emission suggests that both are the result of the processing of intrinsic emission from the corona and accretion disk by the same structure. We explore the resulting implications on the underlying distribution of covering fraction for all AGNs,  by generating mock observables for the reflection parameter and IR luminosity ratio using empirical relations found for the covering fraction with each quantity. We find that the observed distributions of the reflection parameter and IR-to-X-ray ratio are reproduced with broad distributions centered around covering fractions of at least $\sim40\%-50$\%, whereas narrower distributions  match our observations only when centered around covering fractions of $\sim70\%-80$\%.  Our results are consistent with both independent estimates of the covering fractions of individual objects and the typical covering fraction obtained on the basis of obscured fractions for samples of AGNs. These results suggest that the level of reprocessing in AGNs, including X-ray reflection, is related in a relatively straightforward way to the geometry of the obscuring material.
\end{abstract}


 \section{Introduction}

Understanding the role of supermassive black holes (SMBHs) in the evolution of galaxies remains one of the pressing open questions in astronomy (e.g., \citealt{kormendy13, hickox18}). There exist a number of lines of evidence supporting the coevolution of SMBHs and their host galaxies (e.g., \citealt{bell04, gultekin09, oppenheimer10, mcconnell13}). Most of  an SMBH's growth is thought to occur during its active phases  (e.g., \citealt{marconi04, merloni08,  alexander12}). Furthermore, active galactic nuclei (AGNs) provide the best stage for studying all but the most local SMBHs (e.g., d$\lesssim50$\,Mpc, \citealt{xie17}), because it is during these phases that the nuclear regions emit the most radiation due to  larger rates of gas accretion.

AGNs emit across most of the electromagnetic spectrum with a significant portion of the emission in the infrared (IR) at 1--100\,$\mu$m (e.g., \citealt{antonucci93, efstathiou95, elitzur08, padovani17}). The IR emission is thought to be due to a dusty ``torus'' (e.g., \citealt{krolik86, netzer15}), which is primarily heated as a result of the absorption of the optical and ultraviolet (UV) emission from the accretion disk.  At X-ray energies, including in the 3--79\,keV range probed by the {\em Nuclear Spectroscopic Telescope Array} ({\em NuSTAR}; \citealt{harrison13}), the observed emission is due primarily to the corona above the disk.  This wavelength regime, therefore, provides a window into the intrinsic emission very near to the AGNs, in part seen in the tight relationship that has been found between coronal and disk emission (e.g., \citealt{steffen06, lusso17}). Therefore, we might expect to also find a relationship between reprocessed UV emission, captured by thermal IR emission, and  X-ray emission reprocessed primarily via absorption and reflection (e.g., \citealt{guilbert88, madau93, matt94}). The main spectral signatures of reflection include both a hump in the 10--30\,keV range due to Compton scattering and the Fe\,K$\alpha$ line (e.g., \citealt{george91}), whose narrow core peaking at 6.4\,keV provides strong evidence for interaction with cold material (e.g., \citealt{nandra97, reeves03, levenson06}). 

Together, the X-ray and IR emission allow us to probe the nature of the structure that reprocesses nuclear emission.  In particular, the degree of clumpiness in this structure (e.g., \citealt{fritz06, nenkova08}), the relation of its properties to the AGN luminosity, and the distribution of its covering fraction for the AGN population are among the aspects of this structure that are still not completely understood. This last aspect is still poorly constrained both for all AGNs and for only the subset of obscured AGNs and typically examined using complex spectral models (e.g., \citealt{murphy09}). \citet{yaqoob11} examined the dependence of the IR-to-X-ray luminosity ratio on other model parameters in one of these torus models, specifically MYTorus. They found that the ratio was relatively insensitive to column density and instead depended much more strongly on covering fraction and shape of the X-ray continuum. 

Previous studies have found tight correlations between continuum mid-IR (MIR) and intrinsic soft X-ray ($<10$\,keV) luminosities of AGN (e.g., \citealt{lutz04, gandhi09, asmus11, asmus15, chen17}). The absence of a dependence on obscuring column depth in these relations does not meet the expectations of the classical torus models (\citealt{pier93}). These classical models assume smooth and homogeneous dust distributions and  predict a higher amount of obscuration for higher inclinations, resulting in an expected dependence of the reprocessed-to-intrinsic emission ratio on the obscuring column. In contrast, clumpy torus models invoke highly inhomogeneous gas, allowing for unobscured lines of sight even in edge-on configurations  (e.g., \citealt{nikutta09, elitzur12, stalevski16}).  As a result, the clumpiness dilutes the dependence of the reprocessed-to-intrinsic emission ratio on orientation (e.g., \citealt{nenkova08, honig10}).

The {\em Swift}/Burst Alert Telescope (BAT; \citealt{swiftbat05, swift04})  on the {\em Neil Gehrels Swift Observatory},  operating at 14--195\,keV, created the most sensitive hard X-ray survey of the entire sky. Its high energy range is well suited for penetrating large obscuring columns to detect AGNs with very little contamination from other host galaxy emission mechanisms (e.g., \citealt{koss16}). The soft X-ray properties of \sbat AGN have been studied in detail by several studies (e.g., \citealt{winter09, ricci17}). Recently, large subsets from the \sbat\, 58 month and 70 month AGN catalogs have been observed and analyzed separately in the near-IR (NIR), MIR, and far-IR (FIR; \citealt{lamperti17, melendez14, shimizu16, shimizu17}, hereafter S17)  and with detailed hard X-ray spectra taken by {\em NuSTAR} (Balokovi\'{c} et al.  2018, in preparation, hereafter B18).  

Most analyses to date that have jointly used IR and hard X-ray observations of this unbiased sample of local AGNs have primarily explored the connections of the total observed NIR, MIR, and FIR emission, colors, and emission-line properties to the hard X-ray luminosities (e.g., \citealt{mushotzky08, diamond09, rigby09, vasudevan10, matsuta12, ichikawa12, ichikawa17}). However, a joint analysis using a detailed decomposition of the IR spectral energy distribution (SED) combined with good quality spectra extending into the hard X-ray regime has not yet been done for such samples of AGN.\footnote{This type of analysis has been performed for individual objects (e.g., \citealt{farrah16}).}  In this article, we combine the SED decompositions performed by \citet[][S17]{shimizu17} with the {\em NuSTAR} spectral analyses of Balokovi\'{c} et al. (2018, in preparation, B18) of obscured AGNs to constrain the structure of the torus from purely phenomenological modeling. Our sample is one of the largest sample of obscured AGNs with detailed determination of their IR and hard X-ray properties. \\

This article is organized as follows. In Section 2, we describe the sample selection, followed by a summary of the data reduction and parameter extraction undertaken (Section 3). In Section 4, we discuss our analysis  and modeling, as well as the resulting implications, and we summarize our conclusions in Section 5. Throughout this article, we assume a cosmology with Hubble constant $H_{0}$ = 70 km\,s$^{-1}$\,Mpc$^{-1}$, matter density parameter $\Omega_{\rm M} = 0.3$, and dark energy density $\Omega_{\Lambda} = 0.7$ (\citealt{spergel07}). Unless otherwise specifically stated, uncertainties are 1$\sigma$ errors.

\section{Sample}

The sample presented in this work is the overlap of two other subsamples of the \sbat 58 month and 70 month catalogs$^{2, 3}$ (\citealt{swiftbat70}), specifically the \herschel sample of S17 and the \nustar sample of B18.  The S17 sample is composed of  313 \sbat AGNs at z$\,<\,$0.05 that are not blazars or BL Lac objects selected from the 58 month \sbat catalog.\footnote{\href{https://swift.gsfc.nasa.gov/results/bs58mon}{https://swift.gsfc.nasa.gov/results/bs58mon}} It contains an approximately even mix of Seyfert types, based on optical spectra. A small fraction of S17 sample ($<$5\%) are unclassified AGNs or have low ionization nuclear emission-line region (LINER) nuclei. The entire sample was observed  with \herschel in five bands at 70$\mu$m, 160$\mu$m, 250$\mu$m, 350$\mu$m, and 500$\mu$m.

We cross-correlated the S17 sample with the subset of the B18 sample at z$\,<\,$0.05, which contains 95 AGNs selected from the 70 month \sbat catalog\footnote{\href{https://swift.gsfc.nasa.gov/results/bs70mon}{https://swift.gsfc.nasa.gov/results/bs70mon}} to have 14--195\,keV flux greater than $1\times10^{-11}$\,erg\,s$^{-1}$\,cm$^{-2}$ and be identified as a narrow-line AGN (i.e., Sy1.8, Sy1.9, or Sy2)\footnote{There has been some work suggesting that late intermediate Seyfert types (Sy1.8, Sy1.9) are more similar to unobscured (e.g., \citealt{sternj12, hernandez17}), although \citet{koss17} recently showed that Sy1.9 AGNs could have column densities up to the Compton-thick regime. Only 12 sources in our sample fall into this category and they are not clustered in the parameters we examine. As such, we do not believe their inclusion biases our conclusions.} in  that catalog. They were all observed  simultaneously with short \nustar and {\em Swift}/X-ray Telescope (XRT; \citealt{swiftxrt05}) observations, typically 20\,ks and 7\,ks respectively. Sources with complex spectra (requiring models with multiple additional components beyond those described in Section 3.1)\footnote{The five AGNs excluded for this reason are the Circinus Galaxy, NGC\,424, NGC\,1068, NGC\,1192, and NGC\,4945.} or low signal-to-noise spectra ($\lesssim$\,300 counts) were also excluded from the B18 sample for greater uniformity in the quality of the X-ray spectral analysis. 

\begin{deluxetable*}{lclcc}
\tabletypesize{\scriptsize}
\tablecaption{Sample Comparisons\label{KS_samples}}
\tablewidth{\textwidth}
\centering
\tablehead{
\multicolumn{3}{c}{Samples} & \colhead{KS Statistic} & \colhead{KS Probability} 
}
\startdata
z$\,<\,$0.05 BAT AGN (gray) & versus & S17 ({\em Herschel}; yellow) & 0.044 & 90.5\% \\
z$\,<\,$0.05 BAT AGN (gray) & versus & z$\,<\,$0.05 B18 ({\em NuSTAR}; green) & 0.109 & 32.1\% \\	
z$\,<\,$0.05 BAT AGN (gray) & versus & this paper (blue) & 0.103 & 56.0\% \\	
S17 ({\em Herschel}; yellow) & versus & this paper (blue) & 0.130 & 29.0\% \\
z$\,<\,$0.05 B18 ({\em NuSTAR}; green) & versus & this paper (blue) & 0.055 & ~99.96\% \\
z$\,<\,$0.05 BAT Sy2 AGN$^{\dagger}$ & versus & this paper (blue) & 0.080 & 90.6\% 	
\enddata
\tablecomments{ Results of performing Kolmogorov--Smirnov (KS) tests on the distributions shown in Figure \ref{sbat}. For ease of comparison to the figure, we note the associated color of the distribution in Columns 1 and 2. The BAT AGN samples are selected from the 70 month catalog. Column 3 has the  KS statistic, corresponding to the largest separation between the cumulative distribution functions of the two samples. Column 4 has the associated probability of the null hypothesis that the two samples originate from the same parent population. We require a probability less than 0.3\% to reject the null hypothesis at a 3$\sigma$ confidence level.\\
$^{\dagger}$ This subset is not explicitly shown in Figure \ref{sbat}.}
\end{deluxetable*}

\begin{figure}
\centerline{\includegraphics[trim=0.2cm .1cm 0.2cm .5cm, clip, width=\linewidth]{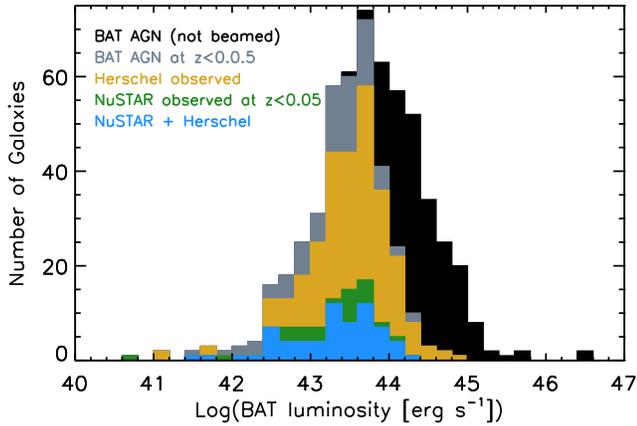}}
\caption{Histogram of the 14--195\,keV \sbat luminosity of all BAT AGNs (black)  from the 70 month \sbat catalog with the exclusion of beamed sources, as well as its subset after a redshift cut at z$\,=\,$0.05 (gray) compared to the samples observed with \herschel (yellow; S17) and \nustar (green; B18 with a z$\,=\,$0.05 redshift cut). The overlap sample that we use is shown in blue. Within the redshift range of z$\,<\,$0.05, the \nustar and \herschel samples are statistically representative of the BAT AGN, as is our joint sample (see Table \ref{KS_samples}).}
\label{sbat}
\end{figure}

There are 69 AGNs in common to these two samples, which we use for our analysis in this work. Their names and coordinates are given in Table \ref{irpar}. This sample is one of the largest of obscured AGN with this high quality of IR and hard X-ray data. Given the varied selection criteria of the S17 and B18 samples and our combination thereof, we investigated how well each of them, as well as our overlap sample of 69, is representative of the full \sbat AGN sample  (excluding beamed sources) and of the \sbat Sy2 AGN sample. As shown in Figure \ref{sbat}, we compared the distributions of the 14--195\,keV \sbat luminosities from the 70 month catalog (\citealt{swiftbat70}) for  all unbeamed BAT AGN at z$\,<\,$0.05 (gray), the full S17 sample (yellow), the z$\,<\,$0.05 B18 subsample (green), and our overlap sample of 69 (blue). Table \ref{KS_samples} contains the results of Kolmogorov--Smirnov (KS)\footnote{Using the {\sc idl} routine {\sc kstwo}.} tests on these distributions, as well as the comparison of the full Sy2 subset of the z$\,<\,$0.05 BAT AGN sample. In each comparison, we cannot reject the null hypothesis that the two samples are consistent with originating from the same population. As a result, we consider our sample to be representative of the complete \sbat AGN sample at redshifts z$<0.05$.

\section{OBSERVATIONS and DATA PROCESSING}
\subsection{NuSTAR and X-Ray Spectral Analysis}

Detailed discussion of the \nustar analysis can be found in B18. We briefly summarize it here. The reduced spectra were binned to have constant signal-to-noise ratios in each energy bin. Each spectrum is fit in the full \nustar energy band (3--79\,keV) in combination with the {\em Swift}/XRT data (0.2--10\,keV) with {\sc Xspec} (\citealt{xspec}). The model used\footnote{const$\times$phabs$\times$(zphabs$\times$cabs$\times$cutoffpl + const$\times$cutoffpl + pexrav + zgauss).} is composed of several components behind an obscuration screen due to foreground absorption by the Milky Way: (1) an absorbed, exponentially cutoff power law for the underlying intrinsic emission; (2) an unabsorbed exponentially cutoff power law to account for the soft emission that may be due to optically thin scattering, X-ray binaries, and/or other ionized emission within the galaxy;  and (3) a reflection component using just the reflection part of the \texttt{pexrav} (\citealt{Magdziarz95}) model combined with an unresolved ($\sigma=10^{-3}$\,keV) Gaussian Fe\,K$\alpha$ line at a fixed rest frame energy of 6.4\,keV. The unabsorbed power law is primarily constrained by the {\em Swift}/XRT data, which is not sufficient to independently constrain the slope, so it is assumed to be the same as that of the intrinsic power law. High energy cutoffs are fixed at 300\,keV, which was justified post facto (B18). 

In the \texttt{pexrav} model, the reflection parameter is restricted to be below zero (i.e., the range in which only the reflection component appears), and a solar metallicity and an inclination of the default 60$^{\circ}$ are assumed.\footnote{\label{ftn:RpexInc} There is a degeneracy in \texttt{pexrav} between inclination and the normalization of the reflection component (e.g., Fig. 1 in \citealt{dauser16}).  However, changes in the inclination have very little effect on the shape of the spectrum. We fix the inclination to handle the normalization only through the reflection parameter $R_{\rm pex}$.} Although the \texttt{pexrav} model is less physically motivated than some more complex models for reflection (see, for example,  \citealt{gandhi14}, \citealt{annuar15}, and \citealt{balokovic14, balokovic18_model} for comparisons between such models and \texttt{pexrav}-based modeling), it has the benefit of capturing the general nature of the reflection with the fewest possible parameters.  A detailed systematic comparison of \texttt{pexrav} and geometrically motivated torus models for a large sample of 120 AGNs, including those used in this work, is in preparation (Balokovi\'{c} et al. 2018, in preparation), with some preliminary results outlined in \citet{balokovicPHD}.

The resulting fits yield the following parameters: the obscuration column density $N_{\rm H}$, the power law slope $\Gamma$,  the equivalent width of the Fe\,K$\alpha$ line (EW$_{\rm Fe\,K\alpha}$), the relative normalization of the unabsorbed power law, and the reflection parameter from the \texttt{pexrav} model ($R_{\rm pex}$=$|R|$, where $R$ is the negative number from the {\sc Xspec} fitting). These parameters are given in Table \ref{nustar}.  In addition to the luminosities described below, we primarily use the reflection parameter and the column density in the analysis that follows, although  Appendix \ref{sec:XIRrel}
  contains additional discussion of the other X-ray parameters.

For three AGNs, the quality of the spectra was insufficient to robustly fit all the parameters, so we fixed one of the parameters: for ESO\,005-G004, we fixed the power law slope at the typical AGN slope of 1.8 (\citealt{Piconcelli05, dadina08}); for MCG+04-48-002, $\Gamma=1.8$ produced an unstable fit, so we used a fixed $\Gamma=1.7$ instead; for the Compton-thick source LEDA96373, the simple model fit only stabilized if the column was fixed, so we used a log($N_{\rm H}$\,[cm$^{-2}$])=24.1, which has been confirmed as reasonable using more complex models (see B18 for further details). For one determination of $N_{\rm H}$, seven of Fe\,K$\alpha$ and nine of the reflection parameter, there is a best fitted value but the lower limit on its uncertainty is poorly constrained.

For our analysis, we use both the spectral parameters described above and the observed,  reflected, and intrinsic 10--50\,keV luminosities. The intrinsic luminosity is corrected for both absorption (which decreases observed flux) and reflection (which increases observed flux) and is, therefore, smaller than the unabsorbed luminosity, which is only corrected for obscuration. These are given in Table \ref{lumin}.  The reflected luminosity is given by the reflection parameter times the intrinsic luminosity.

We chose to use the 10--50\,keV luminosity for this analysis, but we also tested the analysis we undertook using the intrinsic 2--10\,keV luminosity. We found very similar results, because the intrinsic luminosities in both bands are calculated using the same power law model, in which the photon index\footnote{Defined such that $P_E{\rm[ photons\,s^{-1}\,keV^{-1}]}\propto E^{-\Gamma}$.}, $\Gamma$, relates to flux density with $F_{\nu}\propto \nu^{-\Gamma+1}$. The range of $\Gamma$ in our sample introduces $\sim$0.1\,dex of scatter in the intrinsic X-ray luminosity; however, this is relatively small compared to the 0.4\,dex scatter in the ratio between IR and X-ray emission, as we discuss further below.

\subsection{Herschel and Infrared \\Spectral Energy Distribution Fitting}

\citet{melendez14} and \citet{shimizu16} describe in detail the \herschel observations of 313 \sbat galaxies taken by the Photodetector Array Camera and Spectrometer (PACS; \citealt{poglitsch10}) and Spectral and Photometric Imaging Receiver (SPIRE; \citealt{griffin10}), respectively, as well as their reduction and analysis. PACS observations were taken at 70\um and 160\um, whereas SPIRE observations were taken at 250\um, 350\um, and 500\um, all primarily as part of a Cycle 1 program (OT1\_rmushot\_1; PI R. Mushotzky). We briefly summarize the SED analysis done with them below.

\begin{figure}[t]
\centerline{\includegraphics[trim=0.2cm .8cm 0.2cm 0cm, clip, width = \linewidth]{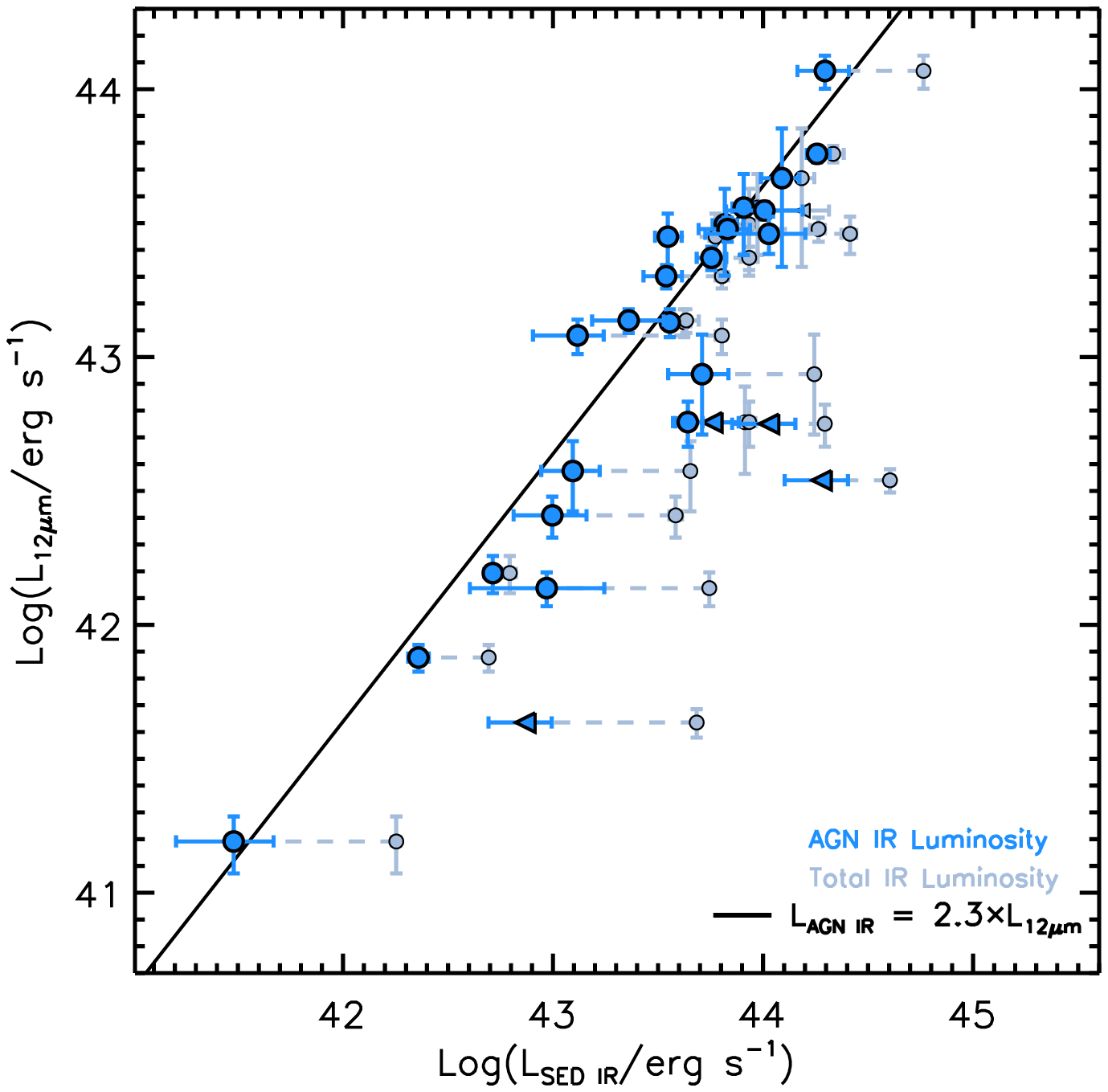}}
\caption{AGN IR (8--1000\,$\mu$m) luminosity  in blue from the SED decomposition compared to  resolved nuclear 12\,$\mu$m luminosities from \citet{asmus14} where available.  We also show the total IR luminosity before the decomposition (gray) and the expected relation (solid line) between the 8--1000$\mu$m luminosity and the 12\,$\mu$m luminosity from the AGN SED models of \citet{mullaney11}. Our AGN IR luminosities  typically agree within their uncertainties with this expected relation, demonstrating the reliability of the SED decomposition compared to high-spatial-resolution MIR observations. Triangles indicate 3$\sigma$ upper limits in a direction of the point.}
\vspace{1mm}
\label{comp12}
\end{figure}

S17 combined these data with archival {\em Wide-field Infrared Survey Explorer} (\wise; \citealt{wright10}) photometry from 3.4 to 22\um to create and fit SEDs and determine the relative contributions of the AGN and star formation (SF) to the IR SED. They model the SEDs as the combination of a modified blackbody (where the dust emissivity inversely depends on wavelength to the power $\beta=2$) and an exponentially cutoff power law (with a fitted power $\alpha$) with turnover wavelength ($\lambda_{\rm C}$). The fitting is done within a Bayesian framework with a Markov chain Monte Carlo to determine the posterior probability distribution functions of the parameters. Through identical analysis of the Herschel Reference Survey (HRS; \citealt{boselli10}), a sample of galaxies that contain only low-luminosity AGNs if any, S17 showed that a component of the power law emission was due to  SF. They used this HRS analysis to determine the correction needed according to the luminosity of the modified blackbody component, which is strictly due to SF.

As a means of testing this SED decomposition, we cross-correlated our sample with that of \citet{asmus14}, who performed high-spatial-resolution MIR photometry of local AGNs. Our samples have 26 AGNs in common. Figure \ref{comp12} shows that the 12\,$\mu$m luminosities from \citet{asmus14} correlate very well with the SED-derived AGN (8--1000$\mu$m) IR luminosities, with scatter about a factor of 3 lower than that of the IR luminosities before the decomposition. The comparison line  shown assumes the ratio between 12\,$\mu$m and the broadband AGN IR luminosities from \citet{mullaney11}.  The relation between the AGN IR and resolved 12\,$\mu$m luminosities of our sample typically agrees with this ratio within the uncertainties of the measurements.

For our analysis, we  primarily used the derived AGN IR luminosity.  We also examined the SF IR luminosity, total IR luminosity (8--1000\um), the AGN luminosity fraction, and the two parameters from the power law (AGN) component ($\alpha$ and $\lambda_{\rm C}$). These parameters are given in Tables \ref{irpar} and \ref{lumin}. For four AGNs, the AGN IR luminosity is a lower limit, likewise restricted by an upper limit on the total IR luminosity. For our analysis, we assign these AGNs the average luminosity between these limits using the range to the limits as the uncertainty on these values.

\begin{figure*}[t]
\centerline{\includegraphics[trim=0.9cm .7cm 0.1cm 0.2cm, clip, width=\linewidth]{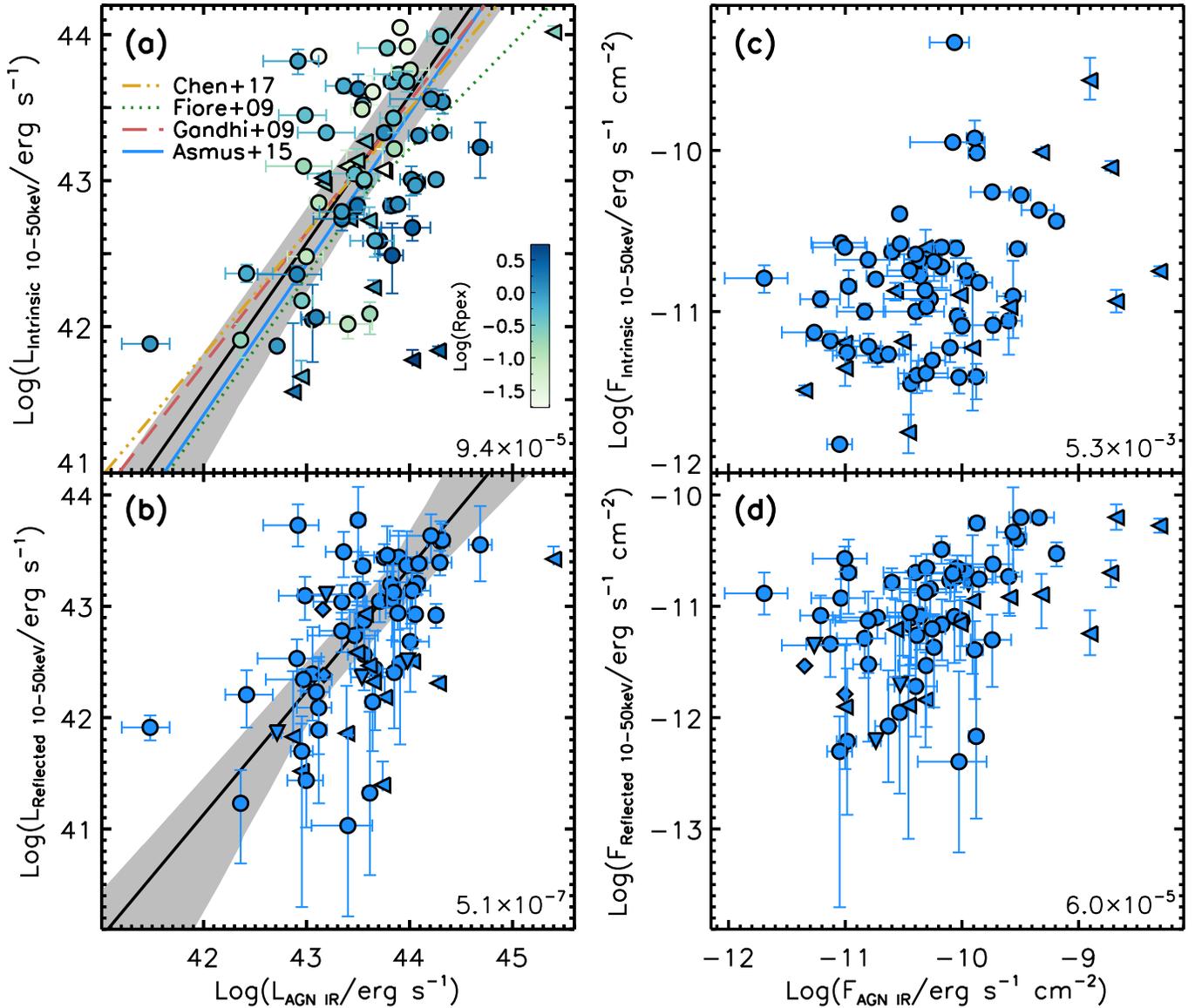}}
\caption{AGN IR  (8--1000\,$\mu$m) luminosity from the SED decomposition compared to intrinsic AGN 10--50\,keV {{(a)}} and reflected 10--50\,keV luminosities {{(b)}},  as well as the corresponding plots using fluxes {{(c, d)}}. Triangles indicate 3$\sigma$ upper limits in a direction of the point; diamonds are upper limits in both directions. In panel {{(a)}}, the points are color-coded by the logarithm of reflection parameter (or its 3$\sigma$ upper limit), and the literature relations (\citealt{chen17} in yellow dash-triple dotted, \citealt{fiore09} in green dotted, \citealt{gandhi09} in red dashed, and \citealt{asmus15} in solid blue) have been adjusted from their monochromatic IR and 2--10\,keV luminosities using conversion factors from \citet[IR]{mullaney11} and assuming a power law with $\Gamma=1.8$ (X-ray).  The black solid line (surrounded by the gray-shaded region of 3$\sigma$ confidence,  derived from a bootstrapping analysis) is the best linear fit to the data (see text).  The correlation between these luminosities is also seen between the fluxes; the probability of not having a correlation is given in the lower right.}
\label{hi_rp}
\end{figure*}

\section{ANALYSIS AND DISCUSSION}
\label{sec:disc}

We explored the relations between and within the IR and X-ray properties, including (1) {\em NuSTAR} spectral parameters,;(2) IR modeling parameters; (3) intrinsic,  reflected, and observed X-ray luminosities in the 10--50\,keV  band; (4) AGN, total, and SF IR luminosities; and (5) ratios of an IR luminosity to an X-ray luminosity. We show a subset in Figures  \ref{hi_rp} and \ref{hi_nh}.  For each of these pairings, we use the \texttt{ASURV} survival analysis package to calculate the Spearman $\rho$ rank correlation (\citealt{asurv92, asurvCorr}), thereby taking the limits into account. This statistic tests the null hypothesis that there is no monotonic relation between the parameters. We define a significant correlation as one  that rejects this hypothesis by having a probability less than $3\times10^{-3}$ {(log[$p{]\le-2.52}$)}, corresponding to approximately 3$\sigma$. To calculate the confidence interval of the Spearman statistic and associated probability, we undertook a bootstrap analysis in which we pick 1000 samples and ran the \texttt{ASURV} analysis on each.\footnote{The code we wrote to do this analysis is available at \\ \url{https://github.com/lalanz/bootstrap_asurv}.}

In the sections below, we discuss in detail how the correlation between reflected X-ray and IR emission implies a common source of reprocessing of the intrinsic emission and the implications of the relationship between the reflection parameter and the ratio of IR-to-X-ray emission for the distribution of covering fractions for all AGN.  Appendix \ref{sec:XIRrel} contains additional discussion of the relations of other X-ray and IR parameters.

\subsection{Relationship Between IR and X-Ray\\ Intrinsic and Reflected Luminosities}
\label{sec:lumin}

We begin by comparing X-ray intrinsic and reflected luminosities to the IR luminosity of the AGN. Correlations between intrinsic X-ray and IR luminosities have long been known, and we show four X-ray to MIR literature relations in Figure \ref{hi_rp}a (\citealt{asmus15}, solid blue; \citealt{chen17}, yellow dash-triple dotted; \citealt{gandhi09}, red long-dashed; and \citealt{fiore09}, green dotted), adapted to account for different IR and X-ray bands. Specifically, the Fiore and Chen relations were derived for IR luminosity at 6\um, whereas the Gandhi and Asmus relations are calculated at 12\um. We convert the relations to the 8--1000\,$\mu$m IR luminosity measured by S17 using the typical ratios provided by \citet{mullaney11}. Similarly, the four relations are derived for X-ray luminosities in the 2--10\,keV band. We convert to the 10--50\,keV band, assuming a power law with $\Gamma=1.8$ (\citealt{netzer15}; consistent with our median $\Gamma$), resulting in a multiplicative factor of 1.38. On these relations, we overlay the AGN's IR luminosity (from S17) against the intrinsic 10--50\,keV luminosity from the fits by B18. 

We find a correlation between these luminosities (Fig. \ref{hi_rp}a; $\rho=0.47\pm0.10$; {log($p{)=-4.03\pm1.35}$}).\footnote{\label{ftn:asurv} We used bootstrap samples picked with replacement from our data. We found that this methodology yielded a larger confidence interval than selecting samples using Gaussian distributions centered at each detection with widths given by their uncertainties. This difference indicates that the uncertainty in our correlations is primarily driven by the sample size and/or intrinsic scatter. We report the median and confidence interval of the statistic and corresponding probability.} The correlation is less significant but still very suggestive when we use fluxes (Fig. \ref{hi_rp}c; $\rho=0.34\pm0.11$; log($p)=-2.28\pm1.17$) instead of luminosities, which confirms that the correlations are not purely due to those that can be introduced into luminosity correlations by the effects of distance (e.g., \citealt{feigelson83}). 

We also find a significant correlation between the reflected X-ray and IR luminosities ($\rho=0.61\pm0.08$; {log($p{)=-6.29\pm1.51}$} for luminosities in Fig. \ref{hi_rp}b; $\rho=0.49\pm0.11$; {log($p{)=-4.22\pm1.58}$} for fluxes in Fig. \ref{hi_rp}d).  On the basis of these confidence intervals, we find a suggestive difference in the correlations, present in both the luminosities and the fluxes, of $\sim1\sigma$, corresponding to a confidence level of 70\%. Because the size of our confidence intervals is primarily driven by our sample size (see footnote {\ref{ftn:asurv}}), a larger sample will be needed to conclusively determine whether the reflected emission is indeed significantly more correlated than the intrinsic emission. 

We also test this relative correlation using a comparative partial correlation test.\footnote{We used the IDL routine \texttt{p\_correlate} solely with the detected luminosities.} We calculate the correlation of the reflected X-ray luminosity with the residual after the correlation between the intrinsic X-ray luminosity with the IR luminosity has been removed, as well as the correlation when we reverse the roles of the reflected and intrinsic X-ray luminosities. We find that the partial correlation is stronger with the reflected X-ray luminosity ($p_{\rm X_R\,IR\,\cdotp X_I} =0.36$) than with the intrinsic X-ray luminosity ($p_{\rm X_I\,IR\,\cdotp X_R} =0.22$). This difference remains when we uses fluxes instead of luminosities ($p_{\rm X_R\,IR\,\cdotp X_I} =0.29$ vs. $p_{\rm X_I\,IR\,\cdotp X_R} =0.18$).

To investigate the relations between these luminosities, we fit a line in Figures \ref{hi_rp}a and \ref{hi_rp}b.  To take into account uncertainties in both luminosities when fitting each line, we perform a fit using orthogonal regression, maximizing the likelihoods provided in \citet{pihajoki17} for both uncensored and censored\footnote{We exclude points that are simultaneously censored in both luminosities.} data. We use the IDL package \texttt{MPFIT}'s Levenberg--Marquardt algorithm to minimize the inverse of the likelihoods (\citealt{mpfit, more78}). We undertook a bootstrapping analysis using 10,000 samples selected with replacements in order to estimate the uncertainties in the slope and intercept.\footnote{The code we wrote to do the orthogonal fit and estimate its confidence interval is available at \\ \url{https://github.com/lalanz/orthogonal_regression}.}

The black solid lines in Figures \ref{hi_rp}a and \ref{hi_rp}b show the results  with the gray regions showing the 3$\sigma$ confidence range from the bootstrapping analysis. The relation with intrinsic X-ray luminosity has a slope of $1.01\pm0.10$, whereas the relation with the reflected luminosity has a slope of $1.11\pm0.13$. The intrinsic X-ray luminosity relation that we find is also  mostly consistent with the literature relations within our confidence interval even without the additional comparison uncertainty due to differences in fitting methodology. The scatter relative to the fits is about a factor of 2 larger  in Fig. \ref{hi_rp}a (for the correlation with intrinsic $L_{\rm X}$) than in Fig. \ref{hi_rp}b (with reflected $L_{\rm X}$).  

\subsubsection{Implications of the Luminosity Correlations}

These analyses  support the idea that the reflected X-ray and IR emission are more strongly correlated than the intrinsic X-ray and IR emission. The correlation between the offsets from the Type 1 AGN IR--X-ray relations (e.g., \citealt{chen17}) and the reflection parameter (color-scale in Figs. \ref{hi_rp}a; see also Section \ref{sec:irx}) suggests either that obscuration is responsible or that the relation reflects a physical link due to the processes affecting both. However, we do not find a correlation between column density and the 10--50\,keV luminosity (Figure \ref{hi_nh}a; $\rho=-0.024\pm0.124$; log($p)=-0.073^{+0.073}_{-0.441}$), the IR-to-X-ray (intrinsic) ratio (Fig. \ref{hi_nh}b; $\rho=0.12\pm0.12$; log($p)=-0.48^{+0.48}_{-0.66}$), the IR-to-X-ray (reflected) ratio (Fig. \ref{hi_nh}c; $\rho=0.12\pm0.11$;  log($p)=-0.45^{+0.45}_{-0.56}$), or the reflection parameter (Fig. \ref{hi_nh}d; $\rho=0.082\pm0.065$; log($p)=-0.30^{+0.30}_{-0.53}$). Therefore, it is unlikely that the X-ray reflection and IR emission correlation is merely due to the optical depth of  obscuring material.

 This suggests that, on average, both the reflected X-ray emission and IR luminosity have been processed by the same, or at least a closely related structure, classically described as the ``torus,'' although the luminosity relations do not specifically imply a particular geometry. Nuclear luminosity, composed of X-rays from the corona and the tightly related optical/UV emission from the accretion disk (e.g., \citealt{steffen06, lusso17}), will interact with this structure. Some of the X-ray emission will be reflected by gas, and a fraction of the total luminosity (dominated by the optical/UV from the disk) will be absorbed and reprocessed into thermal emission from the dust that we observe in the IR. As a result, the correlations we have found between the emission traced by the reflected X-rays and the accretion luminosity reprocessed into the IR may provide insights into the structure with which the nuclear emission is interacting, as we discuss further below.

Our analysis has one further implication that will require more detailed modeling to fully investigate. The \texttt{pexrav} model of reflection off of an infinite slab implicitly assumes interaction with Compton-thick gas.  This assumption combined with a common structure resulting in both the IR reprocessing and the X-ray reflection has one of two possible implications. Either the IR is due to reprocessing by Compton-thick material or there should be similar relations between parameters expressing the interaction of the nuclear emission with the surrounding Compton-thick and Compton-thin gas components. Because Compton-thin (log($N_{\rm H}\,{\rm [cm^{-2}])\simeq22-24}$) obscuration is typically optically thick to the UV emission, which is then reprocessed into IR emission (e.g., \citealt{fabian08}), the second possibility appears to be the more likely scenario. This scenario implies that tori models that include a two-phase medium containing denser, often Compton-thick, clumps dispersed within a more diffuse medium (e.g., \citealt{nenkova08, honig10, stalevski12, feltre12, siebenmorgen15}) should have similar, or at least correlated, covering fractions for the clumps and diffuse media.

\begin{figure}[t]
\centerline{\includegraphics[trim=0.8cm .9cm 0.2cm 0.5cm, clip, width=\linewidth]{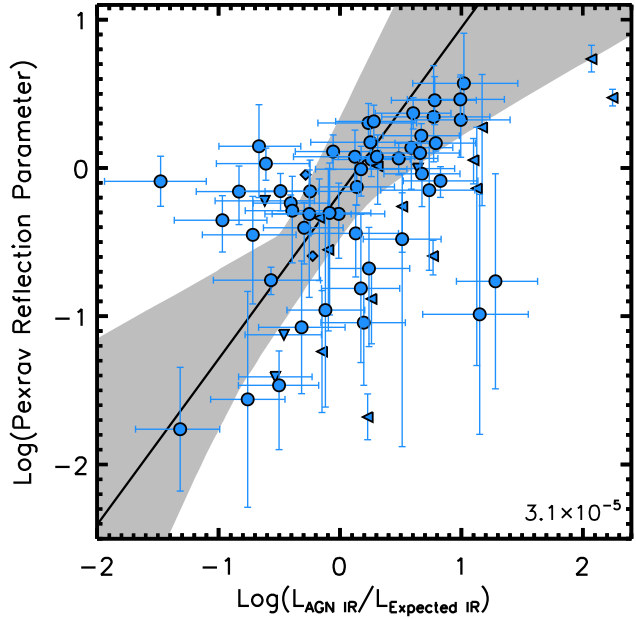}}
\caption{Excess IR luminosity (compared to the expectation from the intrinsic 10--50\,keV luminosity and the Chen et al. (2017) relation) vs. \texttt{pexrav} reflection. Limits (3$\sigma$) in either the IR luminosity (and therefore the IR excess) or the reflection parameter are shown as triangles, unless both are limits, in which case  diamonds without error bars are used. The solid line shows the best fit including the censored data, with the 3$\sigma$ region of confidence  for the fit derived from the bootstrapping analysis delineated by the gray-shaded region. There is a correlation between these parameters, whose  Spearman rank correlation probability  of the absence of a correlation is given in the lower right, which we used to probe the covering fraction distribution. 
\vspace{0.5mm}
}
\label{chen_hi_rp}
\end{figure}

\subsection{Modeling the Distribution of Covering Fractions}

Having found that reflected X-ray luminosity and IR luminosity may both be associated with the same obscuring structure, we investigate the relation between the reflection parameter and the ratio of the intrinsic 10--50 keV luminosity to the IR luminosity and the links of this relation to covering fraction.  Previous studies (e.g., \citealt{yaqoob11}) found that the ratio of IR-to-X-ray luminosities was relatively insensitive to column density. We find a consistent lack of a correlation between the luminosity ratio and column density  (Fig. \ref{hi_nh}b; $\rho=0.12\pm0.12$; log($p)=-0.48^{+0.48}_{-0.66}$) in our purely phenomenological modeling. We therefore investigate the constraints that our modeling imposes on the covering fraction distribution based on the relation between IR-to-X-ray luminosity ratios and the reflection parameter. \\\\\\\\

\subsubsection{Determining the Comparison Parameters: \\Reflection and Infrared Excess}
\label{sec:irx}

In Figure \ref{hi_rp}a, the points are color-coded according to the logarithm of the reflection parameter.  We find a correlation between the intrinsic and reflected X-ray emission ($\rho=0.55\pm0.09$;  log($p)=-5.26\pm1.59$ in luminosities; $\rho=0.40\pm0.11$;  log($p)=-3.02\pm1.45$ in fluxes). This correlation, combined with the relation between reflected X-ray and IR emission, results in a tendency for AGNs in the lower right sector of Fig. \ref{hi_rp}a to have higher reflection parameters. To examine the relationship between reflection and IR emission another way, we calculate the ratio of the observed IR emission compared to the expectation from the \citet{chen17} relation, shown in Figure \ref{hi_rp}a and derived for Type 1 AGNs, to calculate the expected IR emission from the intrinsic hard X-ray luminosity.  We refer to this ratio as the IR excess:

\vspace{-5mm}
\begin{multline}
{\rm log(IR~Excess) = log\Bigg(\frac{Observed~AGN~IR}{Expected~AGN~IR}\Bigg),~where}\\
{\rm log(Expected~AGN~IR) = log(IR_{corr.})+~~~~~~~~~~~~~~} \\
{\rm [log(L_{10-50\,keV} )- C_{1} -  log(X_{corr})]/C_2+45.,}
\end{multline}
in which IR$_{\rm corr.}$ is the \citet{mullaney11} ratio between 6$\mu$m and total IR emission, and X$_{\rm corr.}$ is the ratio between 2--10\,keV and 10--50\,keV luminosity, assuming a $\Gamma=1.8$ power law. Depending on whether the log($L_{\rm 10-50\,keV}$/erg\,s$^{-1}$) is above or below 44.56 (corresponding to a log($L_{6\mu m}$)=44.79), [$C_1$, $C_2$] is [44.51, 0.40] or [44.60, 0.84], respectively (\citealt{chen17}). We plot IR excess against reflection parameter in Figure \ref{chen_hi_rp} ($\rho=0.51\pm0.11$;  log($p)=-4.51\pm1.62$)\footnote{ The strength of this correlation is at a level similar to that between fluxes because the ratio divides out the luminosity distance.}.

We fit this parameter pairing using our orthogonal regression methodology, thereby using the limits and uncertainties on both IR excess and reflection parameter simultaneously. The best-fit line is given by 
\begin{multline}
{\rm log(R}_{\rm pex}) = - (0.17\pm0.21) \\
+ (1.12\pm0.37)\times {\rm log(IR~Excess)}.
\end{multline}
\vspace{-3mm}

\subsubsection{Modeling Observables from Covering Fractions}
\label{sec:model}

\begin{figure*}
\centerline{\includegraphics[trim=0.0cm 0.0cm 0.0cm 0.0cm, clip, width=0.9\linewidth]{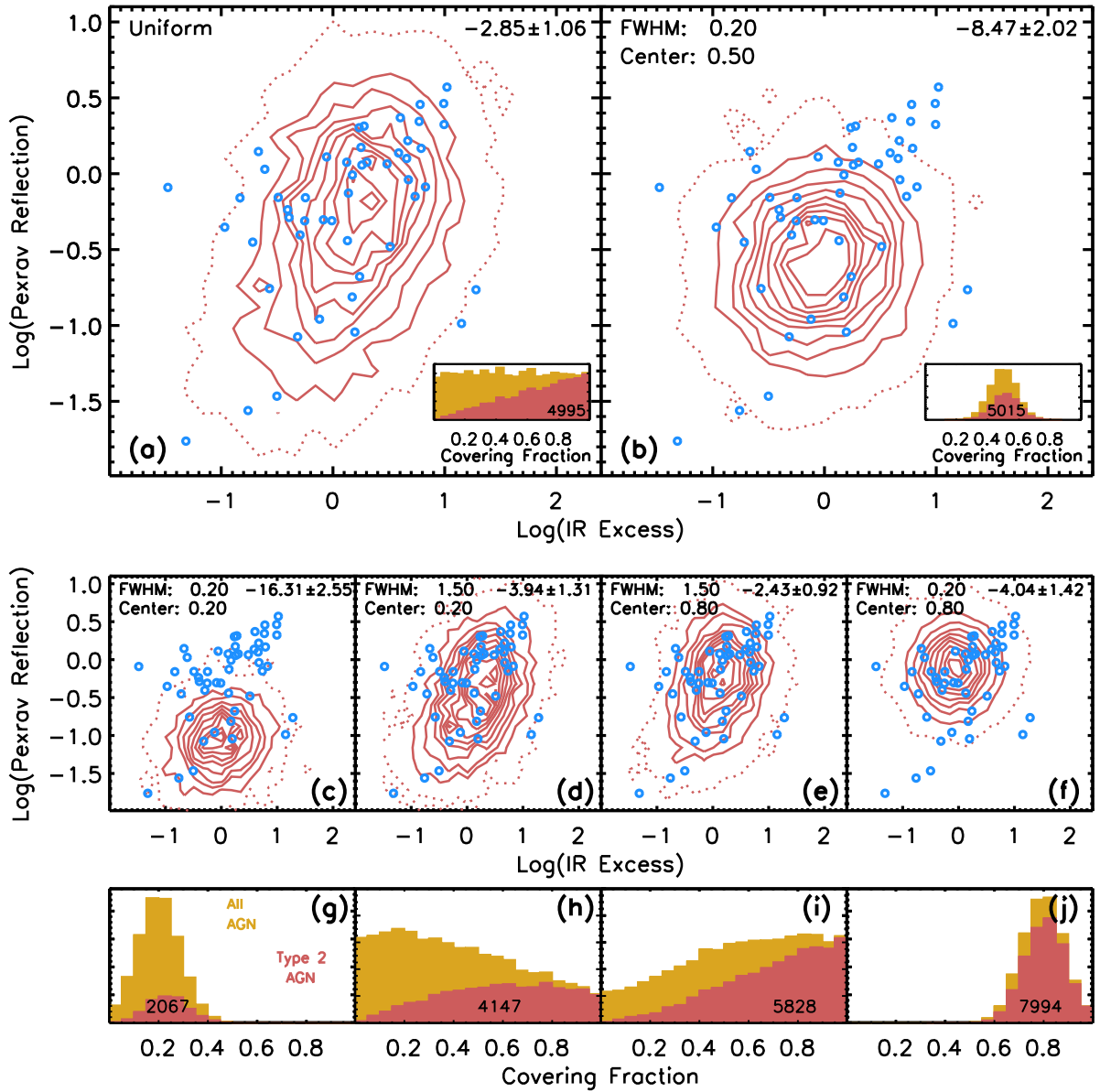}}
\caption{ Contours of mock observables (red: solid at intervals of 10\%, dotted contains 99\%) calculated from modeling undertaken for a range of covering fraction distributions compared to observed \sbat detections (blue circles) shown in Figure \ref{chen_hi_rp}. A brief description of the shape of the distribution is given in the upper left of these panels: {(a)} uniform distribution, {(b)} a narrow centered Gaussian distribution, {(c)} a narrow Gaussian centered at low covering fractions, {(d)} a wide Gaussian centered at low covering fractions, {(e)} a wide Gaussian centered at high covering fractions, and {(f)} a narrow Gaussian centered at high covering fraction. The yellow histograms (insets in {(a)} and {(b)}; {(g)}--{(j)}) show the distribution of covering fraction for the full AGN population, whereas the overlaid red histograms show the distribution for the Type 2 AGN subsample (the number of which is written in black). The values given in the upper right of each panel are the logarithms of the probabilities (and 1$\sigma$ intervals) that the mock observables and observed data have the same two-dimensional parent population (see Fig. \ref{sim_stats} and Section \ref{sec:model_comp}).}
\label{sims_data}
\end{figure*}

We developed a simple model in order to explore the physical origin of this relation, in particular whether we can parameterize it solely on the basis of the covering fraction. With this model, we are implicitly assuming an axisymmetric geometry with relatively constant distribution of the obscuring matter, seen along random lines of sight. We generate  a range of covering fraction distributions of all AGNs, including both wide and narrow Gaussian distributions and a uniform distribution (histograms of Figure \ref{sims_data})  to cover the full range of possible scenarios.  We step through central values of the Gaussian  from 0 to 1 in steps of 0.05 and through values of full width at half maximum (FWHM) from 0.1 to 2 in steps of 0.1, for a total of 420 models. 

For each distribution, we draw 10,000 simulated AGN for which we set the probability of being classified as a Type 2 equal to the covering fraction (e.g., \citealt{elitzur12, netzer15}). We then separate the sample into Type 1 and Type 2 subsamples. For each of the simulated Type 2 AGNs, we calculate a model reflection parameter using:
\begin{equation}
\label{eqn3}
{\rm log(R_{\rm pex}) = 1.7 \times f_{\rm cov} - 1.4},
\end{equation}
 where $f_{\rm cov}$ is the covering fraction (i.e., the fraction of the sky obscured by gas and dust).
This empirical relation is based on the determination of both reflection parameter and covering fraction for a larger sample of \nustar observed AGNs using more complicated models  compared to fits using the phenomenological modeling used in this article (Balokovi\'{c} et al. 2018, in preparation).\footnote{ As discussed in footnote \ref{ftn:RpexInc}, there exists a degeneracy between assumed inclination and $R_{\rm pex}$ in the \texttt{pexrav} model. This empirical relation was derived for values of $R_{\rm pex}$ determined with a fixed inclination of 60$^{\circ}$.} \citet{balokovic18_model} showed the results of this modeling for four galaxies (their Figure 8), which support the linear scaling between $f_{\rm cov}$ and ${\rm log(R_{\rm pex})}$. We also investigate the effect on our conclusions due to the variations on this relation (discussed in Section \ref{sec:robust} and Appendix \ref{sec:altmod}).

Given the correlation between reflected and IR emission, we assume the degree of reflection (and absorption) depends on the covering fraction (e.g., \citealt{maiolino07, treister08, elitzur12}). Therefore, we parameterize ${\rm L_{\rm IR} = \eta \times f_{\rm cov}\times L_{\rm bol}}$. $\eta$ encompasses all other constants of proportionality, including an assumed constant ratio between the optical/UV disk emission and the reprocessed IR emission that is the same for Type 1 and Type 2 AGNs. We use this to determine the relation between the IR excess and the covering fraction. Because our observed IR excess is determined relative to a relation derived for Type 1 AGNs, the intrinsic IR excess is given by the equation:
\begin{equation}
\label{eqn4}
{\rm log(IR~Excess) = log( {\it f}_{\rm cov}\,/<{\it f}_{\rm cov;~Type~1}>) }. 
\end{equation}
The average covering fraction for Type 1 AGNs is calculated from the Type 1 subsample of simulated AGNs.  We investigate the robustness of this parametrization in two ways. First, we relax the assumption of a linear scaling between IR luminosity and covering fraction, using the IR-to-bolometric luminosity dependence on covering fraction of \citet{stalevski16}. Second, we explore the effect of changing the dependence of the expected IR emission (and therefore the IR excess) on the observed intrinsic X-ray emission. These variations and the effect on our conclusions are discussed in Section \ref{sec:robust}  and Appendix \ref{sec:altmod}.

We add scatter to create mock observables for these model values. For the reflection parameter, the magnitude of the scatter is set by the average observed uncertainty of $\sim0.3$\,dex. For the IR excess, we find a scatter in IR-to-X-ray luminosity ratios relative to the literature relations (Fig. \ref{hi_rp}a) of $\sim0.4$\,dex, similar if a bit larger than that found by \citet{gandhi09} in the MIR to X-ray luminosity ratio, as might be expected given the use of SED decomposition in S17 compared to the nuclear MIR fluxes used by \citet{gandhi09}.

\subsubsection{Comparison of Models and Observations}
\label{sec:model_comp}

Figure \ref{sims_data} compares the results of  a subset of our models to our observations. The blue points are the same as those in Figure \ref{chen_hi_rp}, whereas the red contours show the distributions of the mock observables calculated using Equations \ref{eqn3} and \ref{eqn4} from the covering fraction of each simulated Type 2 AGNs. We quantify the likelihood that the observations are consistent with each set of mock observables using a two-dimensional KS (2D-KS) test (\citealt{peacock83,goulding14}). We follow the methodology of \citet{goulding14} and run 10,000 bootstrap samples of the observations and mock observables for which we calculate the 2D-KS statistic and associated probability of the null hypothesis that both samples are consistent with having the same parent population.

\begin{figure*}
\centerline{\includegraphics[trim=0.0cm 0.0cm 0.0cm 0.0cm, clip, width=\linewidth]{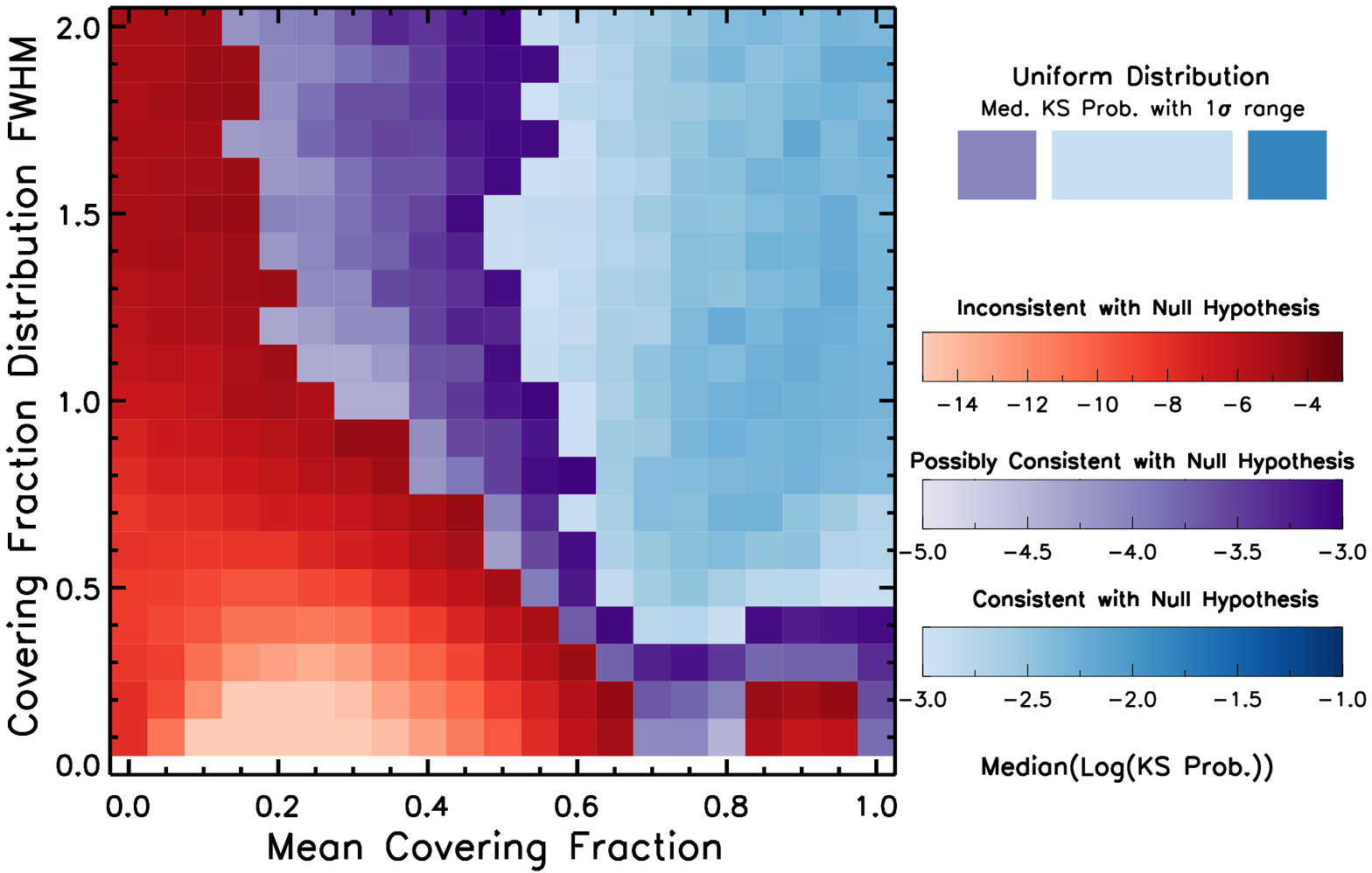}}
\caption{ Distribution of the median 2-D KS probability that the mock observables of the Type 2 AGN subset (i.e., red contours in Fig. \ref{sims_data}) calculated for a Gaussian covering fraction distribution defined by a given FWHM and central value and the observed detections of Figure \ref{chen_hi_rp} are consistent with the null hypothesis of belonging to the same parent distribution. The corresponding probability for a uniform distribution is given in the central color block at the top right with its 1$\sigma$ range shown to the left and right. The three color scales indicate the logarithm of the probability for models where (1) the median probability does not reject the null hypothesis (blue), (2) the median probability significantly rejects the null hypothesis (red), or (3) the median probability is within 1$\sigma$ of not rejecting the null hypothesis (purple; see Section \ref{sec:model_comp}).}
\label{sim_stats}
\end{figure*}

Figure \ref{sim_stats} shows the medians of the probability distribution resulting from each set of bootstrap runs. The three colored blocks at the upper right show the results for the uniform distribution. The larger block in the middle shows that this distribution has a 2D-KS median probability indicative that the null hypothesis cannot be rejected. We find that the probability distributions generated by the bootstrap methodology have a typical breadth of about 1\,dex, as illustrated by the two color blocks to the left and right of the block corresponding to the uniform distribution's median probability. As a result, we use three color scales to indicate the likelihood of the null hypothesis. Models whose median 2D-KS probability is at least $10^{-3}$ are shown with blue colors. For these models, we cannot reject the null hypothesis that the data and mock observables are consistent. Models that significantly reject the null hypothesis by having at least 84\% of their probability distribution (corresponding to all probabilities less than the median+1$\sigma$ probability) less than $10^{-3}$ are shown with red colors. The intermediate set of models, shown in purple, have a median probability less than $10^{-3}$, but the standard deviation of its probability distribution extends above $10^{-3}$. For this set of models, it is possible that the null hypothesis is not rejected, because models shown in very light blue or dark purple have a very similar probability. 

\subsubsection{Implications for Covering Fraction Distributions}
\label{sec:results}

Taken together,  Figures \ref{sims_data} and \ref{sim_stats} provide some insight into the underlying distribution of covering fractions for all AGNs.  Narrow Gaussian models for distributions of covering fractions (e.g., Fig. \ref{sims_data}b, \ref{sims_data}c, or \ref{sims_data}f) tend to poorly match the observations. This is particularly acute for narrow distributions skewed to peak at low covering fractions (e.g., Fig. \ref{sims_data}c). Even very wide distributions skewed to peak at low covering fractions (e.g., Fig. \ref{sims_data}d) at best have only marginal or suggestive indications of agreement with the observations. Broad, centered (e.g., Fig. \ref{sims_data}a) or peaking at high covering fraction (e.g., Fig. \ref{sims_data}e) result in observables that match the data best. We also find that once the distribution has a FWHM of 1.0, a further increase in breadth does not tend to change the degree of agreement. The narrowest distributions whose observables are consistent with the data are centered at covering fractions of $\sim0.70-0.80$.

Broad distributions of covering fraction, combined with the assumption that the likelihood of a Type 2 designation increases with covering fraction (e.g., \citealt{elitzur12}), also have the benefit of yielding distributions of the Type 2 covering fraction similar to what has been observed with more complex modeling. High-spatial-resolution IR studies of small samples of local quasars have found that, although the distributions of covering fractions for Type 1 and 2 AGNs are different, they also overlap significantly (e.g., \citealt{mor09, ramos11, alonso11, ichikawa15}). \citet{mateos16} recently undertook clumpy torus modeling of the NIR to MIR SEDs of 227 X-ray selected AGNs and likewise found broad, overlapping distributions for the covering fractions of both obscured and unobscured AGNs. Their distributions are different from Gaussians or a uniform distribution due to the presence of additional low covering fraction sources.  However, our Type 2 subsets for broad distributions (e.g., Fig. \ref{sims_data}a inset or Fig. \ref{sims_data}i) show similar peaks at high covering fractions and decline with decreasing covering fraction as their Type 2 distribution (red line in their Fig. 3). We used their distribution for all AGNs (black line in their Fig. 3) to generate another set of mock observables. We find that the mock observables from this underlying distribution agree similarly well with our data to some of our very broad models (e.g., Fig. \ref{sims_data}a,  \ref{sims_data}d-e; log($p$)$=-3.53\pm1.24$), indicating that for broad distribution, this analysis is not sensitive to the details of their shape.

Figure \ref{sim_stats} also shows that Gaussian distributions centered at covering fractions of $\sim$0.70--0.80 can have a broad range of FWHM capable of reproducing the observations, down to relatively narrow widths. Interestingly, \citet{ricci15} recently showed that, on the basis of the obscured fractions in the BAT AGN Spectroscopic Survey, the typical covering fraction of AGNs should be approximately 70\%.  Our analysis is consistent with these results, finding that even relatively narrow distributions centered at 70\% yield observables consistent with our data. 

 Despite this consistency with \citet{ricci17} regarding typical covering fractions, our model predicts a different relationship between the strength of the reflection component and $N_{\rm H}$. Specifically, we expect Type 2 AGNs to have stronger reflection, because our modeling tends to give them higher covering fractions. Although we do not find a significant correlation between $R_{\rm pex}$ and $N_{\rm H}$, the median and average $R_{\rm pex}$ of our AGNs with $N_{\rm H} \geq 10^{23}$\,cm$^{-2}$ are larger than for our AGNs with $N_{\rm H} < 10^{23}$\,cm$^{-2}$. This is more consistent with the results of \citeauthor{ricci11} (\citeyear{ricci11}; see also \citealt{vasudevan13, esposito16}). One possible explanation for the closer similarity to the \citet{ricci11} results compared to the \citet{ricci17} results may lie in a modeling degeneracy. In both our modeling and the \citet{ricci11} modeling, the typical $\Gamma$ is consistent across different bins of $N_{\rm H}$. However, in \citet{ricci17}, there is a significant difference in the distribution of the photon indices of the most obscured AGNs compared to the distribution for their less obscured AGNs. Because there exists a degeneracy between $\Gamma$ and the reflection parameter (e.g., \citealt{delmoro17}; see also Appendix \ref{sec:XIRrel}), some of the effect seen in \citet{ricci17} may therefore be induced by the difference in $\Gamma$. 

\subsubsection{Robustness of the Modeling Results}
\label{sec:robust}

Given the phenomenological nature of the spectral modeling used in our analysis, we chose to use a simple model for our  mock observables to limit the number of free parameters. As a result, there are multiple additional considerations that could be taken into account to further constrain the nature of the covering fraction distribution. For example, our model does not include obscuration or reflection due to dust in the polar regions (e.g., \citealt{honig12}, \citealt{lopez16}). Although our model does not assume a geometry that precludes its presence, it may have a different heating mechanism that would not be captured in our model. Additionally, given the relatively small dynamic range of our luminosities, we also do not include a dependence of the covering fraction on AGN luminosity, the so-called receding torus models (e.g., \citealt{lawrence91, simpson05}), although recent results suggest that covering fraction may not vary significantly with luminosity (\citealt{mateos17, ichikawa18}).

 As was mentioned in \ref{sec:model}, the underlying uncertainty in determining covering fraction, including its dependence on other AGN properties, manifests in uncertainty in the empirical relations we use to calculate the mock observables of reflection parameter and IR excess. We explored two variations in each parameter to explore the robustness of our conclusions.  Appendix \ref{sec:altmod} contains a detailed discussion of these alternative empirical relations for the reflection parameter as well as IR excess. The results of these tests are all consistent with our conclusions; specifically, we still find that broad distributions of covering fractions result in mock observables with the best agreement with our data and that the narrowest models yielding observables in agreement with our data tend to be centered around $\sim$70\%. The range of distributions yielding observables consistent with our observations show greater sensitivity to the relation between covering fraction and reflection parameter than to the relation between covering fraction and IR excess.
 
 In determining the reflection parameter for our observations, the inclination of the \texttt{pexrav} was fixed to 60$^{\circ}$, due to the degerenacy between the normalization of the reflection component and the inclination parameter (e.g., \citealt{dauser16}). Inclination does not affect the total IR emission of the torus (e.g., \citealt{stalevski16}), but it can affect the degree to which X-rays are reflected into the observed line of sight. As a result, inclination effects could be responsible for at least some of the scatter in Figure \ref{chen_hi_rp}. However, disentangling this effect will require more complex modeling than that used in this analysis.

\section{Summary}

We performed joint IR and X-ray phenomenological modeling of a large sample of obscured AGNs.  We found a significant correlation between the reflected X-ray and IR emission, with multiple suggestive indications that this correlation is stronger than that between intrinsic X-ray and IR emission. This relation suggests that both the X-ray reflection and the UV emission reprocessed into IR have been processed by the same structure.

We parametrized this effect as a covering fraction, encompassing both geometrical factors and the impact of clumpiness, and investigated which distributions of covering fractions can reproduce the observed distributions of IR excess and reflection parameters. A range of broad covering fraction distributions of the underlying total AGN population (e.g., Fig. \ref{sims_data}a, \ref{sims_data}{e}) results in mock observables, determined from simple empirical relations, consistent with our observations. We also find that the narrowest distributions resulting in observables in agreement with our data are centered around covering fractions of 70\%--80\%. These results are consistent with both other methodologies for estimating covering fraction: the set of independent estimates of the covering fraction of individual objects suggests a broad distribution of covering fraction (e.g., \citealt{mateos16}), and statistical estimates of the typical covering fraction from sample properties (e.g., \citealt{ricci15}) find an expected covering fraction of $\sim$70\%.

Although our modeling was purposely kept simple to investigate how much can be gleaned without the use of complex assumptions, their implications regarding covering fraction distribution are not in agreement with the classical unification model (e.g., \citealt{antonucci93}). In the simplest classical picture, all AGNs have the same covering fraction and opening angle, and it is only orientation that governs whether an AGN is identified as obscured. In contrast, in clumpy torus models (e.g., \citealt{nenkova08, stalevski12}), the covering fraction  depends on the number and distributions of obscuring clouds, possibly embedded in a more diffuse medium. Our  modeling suggests that the clumps and the more diffuse media should have at least correlated covering fraction, but more detailed modeling will be necessary to fully investigate this question. 

\acknowledgements 
 We thank the referee for useful comments, which particularly strengthened the statistical analysis. L.L. and R.\,C.\,H. acknowledge support from NASA through grant number No. NNX15AP24G. L.L. acknowledges support from NASA through grant number NNX17AB58G. R.C.H acknowledges support from the National Science Foundation through CAREER grant number 1554584. M.B. acknowledges support from NASA Headquarters under the NASA Earth and Space Science Fellowship Program, grant No. NNX14AQ07H, and support from the Black Hole Initiative, which is funded by a grant from the John Templeton Foundation. C.R. acknowledges the CONICYT+PAI (Convocatoria Nacional subvencion a instalacion en la academia convocatoria a\~{n}o 2017 PAI77170080), and financial support from FONDECYT 1141218, Basal-CATA PFB--06/2007, and the China-CONICYT fund. F.E.B. acknowledges support from CONICYT-Chile (Basal-CATA PFB-06/2007, FONDECYT Regular 1141218), the Ministry of Economy, Development, and Tourism's Millennium Science Initiative through grant IC120009, awarded to The Millennium Institute of Astrophysics, MAS. M.K. acknowledges support from NASA through ADAP award No. NNH16CT03C. A.M. acknowledges support from the ASI/INAF through grant No. I/037/12/0-011/13. L.Z. acknowledges financial support under ASI/INAF contract I/037/12/0. This work made use of ASURV Rev. 1.2 (\citealt{asurv92}), which implements the methods presented in \citet{asurvCorr}.

{\it Facilities:} \facility{\herschel, ~~{\em NuSTAR}, ~~{\em Swift}, ~~{\em WISE}~.}

\appendix
\renewcommand{\thefigure}{A\arabic{figure}}
\setcounter{figure}{0}

\section{A: Relationships Between Other X-Ray and Infrared Properties} \label{sec:XIRrel}

\begin{figure*}[h] 
\centerline{\includegraphics[trim=0.cm 0.cm 0.cm 0.cm, clip, width=\linewidth]{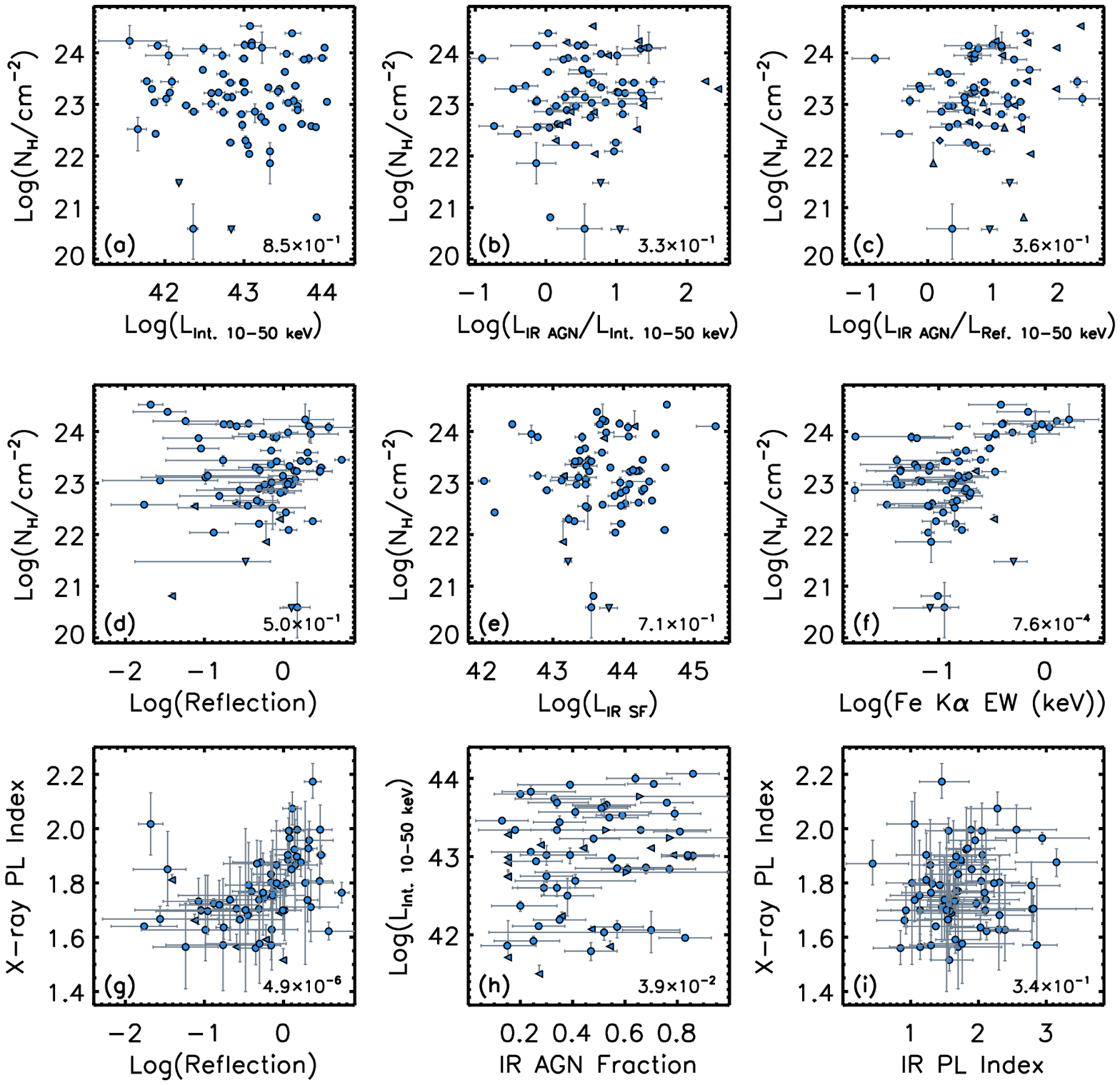}} 
\caption{Each panel shows a comparison of parameters from the X-ray and/or IR fitting. Triangles are 3$\sigma$ limits in the direction of the point. Panel {(a)} shows a lack of correlation between the intrinsic 10--50\,keV luminosity and column density, finding instead a relative consistent range of 1.5--2\,dex of $N_{\rm H}$ derived over the whole range of luminosity. Similarly, we do not find a correlation between column density and the ratio of IR-to-intrinsic X-ray luminosity  {(b)}, the ratio of IR-to-reflected X-ray luminosity  {(c)}, or the reflection parameter  {(d)}, indicating that the relation between X-ray reflection and IR emissions is likely not to be due to obscuration effects. We also do not find a correlation between column density and IR luminosity associated with star formation {(e)}, suggesting that little of the obscuration is due to gas on galactic scales. Panels {(f)} and {(g)} show correlations imposed by the X-ray modeling. The fraction of IR luminosity due to the AGN also does not appear to correlate with the 10--50\,keV intrinsic luminosity {(h)}, and we do not find a correlation between the power law indices in the X-ray and IR fitting {(i)}. The numbers on each plot are the Spearman rank correlation probability of the absence of a correlation.
\label{hi_nh}
}
\end{figure*}

Most pairings of IR and X-ray properties, beyond those discussed in Section \ref{sec:disc} do not yield significant correlations. We show a subset that may be of interest in Figure \ref{hi_nh}. Of those with significant correlations, several are due to definitions of model parameters or model degeneracies. Over the luminosity range of our sample, the anticorrelation of the equivalent width of the ${\rm Fe\,K\alpha}$ line to the observed X-ray luminosities ($\rho=-0.50\pm0.10$; log($p)=-4.42\pm1.76$) is primarily due to the reduction of the continuum level resulting in an increase in equivalent width even at constant line flux. This effect is also seen in the correlation between column density and ${\rm Fe\,K\alpha}$ equivalent width (Fig. \ref{hi_nh}f; $\rho=0.41\pm0.12$;  log($p)=-3.12\pm1.50$), which is due to modeling methodology. As absorption increases, the continuum is depressed but the line flux is not affected, so, as a result, the equivalent width increases. B18 also finds the correlation we identify between the reflection parameter and $\Gamma$ (Fig. \ref{hi_nh}g; $\rho=0.55\pm0.10$;  log($p)=-5.31\pm1.78$) but argue that it is most likely due to model-based degeneracy (see also \citealt{delmoro17}).

Figure \ref{hi_nh}h shows that we do not find a correlation between the intrinsic X-ray luminosity and the dominance of the AGN in the IR ($\rho=0.25\pm0.11$;  log($p)=-1.41\pm0.92$). This lack of a correlation suggests that our sample likely contains a range of galaxy luminosities and, by inference, black hole masses.  This implies a broad range of Eddington ratios (e.g., \citealt{hopkins09}). We also do not find a correlation between the power law indices of the intrinsic X-ray spectrum and its IR counterpart (Fig. \ref{hi_nh}i; $\rho=0.12\pm0.12$;  log($p)=-0.47^{+0.47}_{-0.65}$). However, given that these two power laws trace different emission mechanisms, intrinsic coronal and reprocessed emissions, respectively, the lack of correlation is not unexpected.

 The degree to which galactic-scale dust contributes to the obscuration of AGNs, and the dependence on this relative obscuration on galactic and nuclear properties, remains unclear (e.g., \citealt{rosario12, rovilos12, chen15,  delmoro16, buchner17, ricci17Nat}). We do not find a correlation between $N_{\rm H}$ and the SF IR luminosity from the decomposition (Fig. \ref{hi_nh}e; $\rho=0.045\pm0.121$;  log($p)=-0.15^{+0.15}_{-0.49}$) or with the total IR luminosity ($\rho=0.070\pm0.122$;  log($p)=-0.25^{+0.25}_{-0.57}$) for our sample, indicating it is unlikely that most of the obscuration of our sources is occurring on galactic scales. Given that few of our sources (6 out of 69) have log($N_{\rm H}\,[{\rm cm^{-2}])\leq22}$, we expect that most of our sources will require significant denser obscuration at smaller scales, consistent with the lack of correlations between IR emission associated with SF and $N_{\rm H}$. However, we cannot rule out small contributions to the obscuration from galactic scales.

\section{B: Investigation of Alternative Models} \label{sec:altmod}

\begin{figure}[t]
\centerline{\includegraphics[trim=0.0cm .0cm 0.0cm .0cm, clip, width=\linewidth]{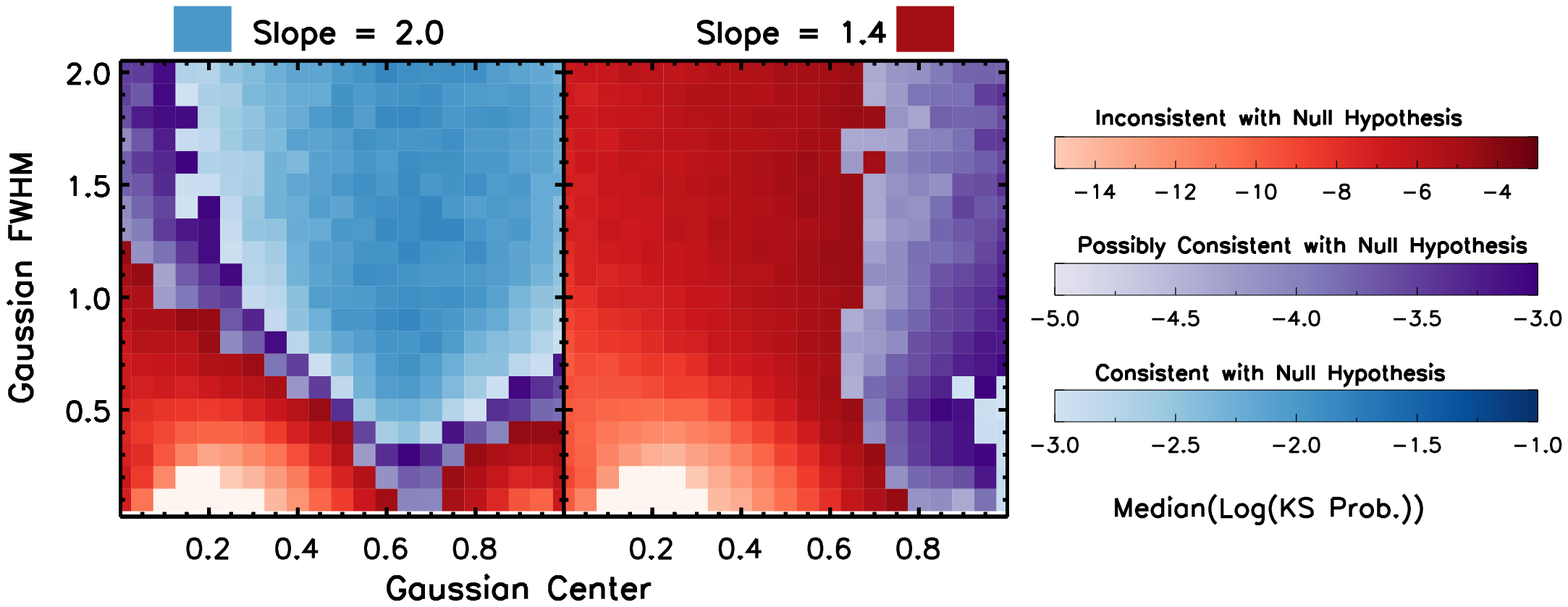}}
\caption{Similar plots to Figure \ref{sim_stats} but with variations in the empirical relation between covering fraction and reflection parameter with steeper (left) and flatter (right) slopes. The colored blocks above each plot are the results for the uniform distribution. In comparison to Figure \ref{sim_stats}, it is clear that the variety of distributions yielding observables consistent with the data depend on the slope, with steeper relations resulting in a larger diversity. However, the conclusions that more types of broader distributions and narrower distributions centered around $\sim$70\%--80\% yield the observables most consistent with the data are still supported by the results with these alternative relations.}
\label{IgR_test}
\end{figure}

 To test the robustness of our conclusions regarding covering fraction distributions, we undertook the same analysis discussed in Sections \ref{sec:model} and \ref{sec:model_comp} for five variations on our analysis. First, we used two different versions of Equation \ref{eqn3}. We opt to retain the simple form of log($R$)$\,\propto f_{\rm cov}$ but investigate the effect on our analysis if we adjust Equation \ref{eqn3}  to have a slope of 1.4 or 2.0, with corresponding intercepts of --1.3 and --1.5, respectively. These variations on Equation \ref{eqn3} are selected to still be consistent with the modeling of Balokovi\'{c} et al. (2018, in preparation) but with more extreme slopes.  Having determined for our original analysis that consistent results were obtained with 1000 or 10,000 bootstrap samples, we ran the 2D-KS analysis using 1000 bootstrap samples for each set of models generated with these altered empirical relations.

Figure \ref{IgR_test} shows the equivalent to Figure \ref{sim_stats} for variations in the empirical relation between covering fraction and reflection parameter. At flatter slopes, the models cover less of the reflection parameter range. As a result, distributions skewed to higher central values will result in better coverage of that parameter and, therefore, those models will agree better with the observations. At steeper slopes, a wider range of models, especially centered at lower covering fractions, agree with our observations. The narrowest models yielding observables similar to the data are still centered around $\sim$70\%. These tests demonstrate that, despite minor changes at the edges of the ranges of models that agree, the conclusion regarding the kinds of models that yield distributions of observables consistent with our data is not very sensitive to the relation between reflection parameter and covering fraction (Eqn. \ref{eqn3}).

\begin{figure}[t]
\centerline{\includegraphics[trim=0.0cm .0cm 0.0cm .0cm, clip, width=\linewidth]{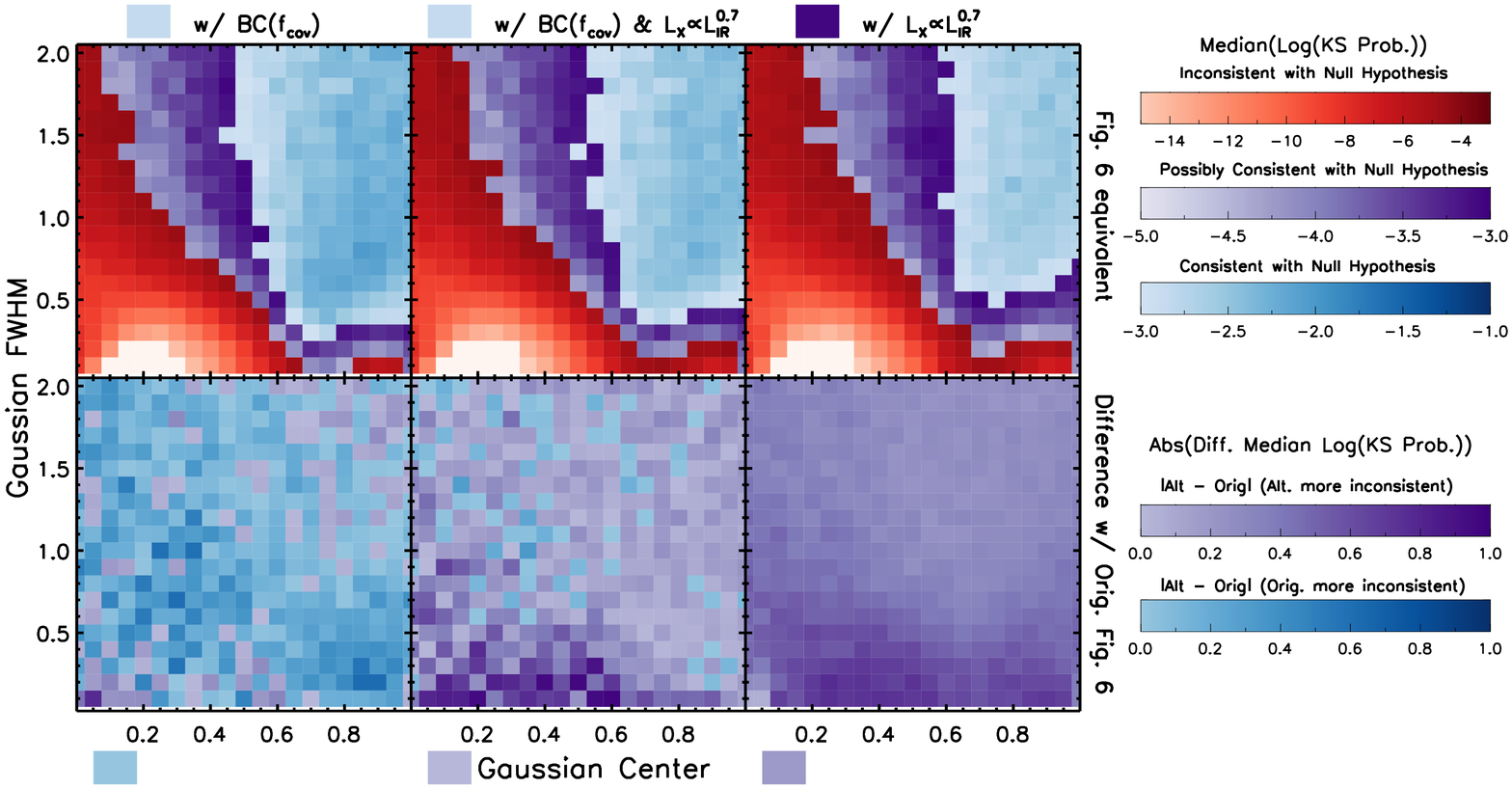}}
\caption{ The results of testing variations in the modeling of IR excess, with the top row showing plots like those shown in Figure \ref{sim_stats} and \ref{IgR_test}. Because the results are so similar to Figure \ref{sim_stats}, in the bottom row, we show the absolute value of the difference with Figure \ref{sim_stats}, using purple to denote when the variation results in greater inconsistency with the null hypothesis and blue when the original model results in greater inconsistency. The left column shows the results with an alternative scaling between L$_{\rm IR}$ and L$_{\rm bol}$ based on Stalevski et al. 2016, the right column shows the results with an alternative calculation of the observed IR excess using a different scaling between  L$_{\rm IR}$ and L$_{\rm X}$, and the middle column includes both changes. In all three cases, the differences are well within the uncertainties of the modeling. The colored blocks above and below each plot show the results for the uniform distribution.}
\label{modIR_test}
\end{figure}

We also examine the impact in variations in the definition of and empirical relation for IR excess. Figure \ref{modIR_test} shows the results of these tests. First, we explored the impact of changing the dependence of the expected IR luminosity on the X-ray luminosity. Instead of using the relation from  \citet{chen17} of ${\rm L_{IR} \propto  L_{X}^{1/0.84}}$, we maintained the assumption that ${\rm L_{IR}\propto L_{bol} \propto L_{UV}}$ and combined it with the relation of the UV emission to the X-ray emission of ${\rm L_{UV} \propto  L_{X}^{1/0.70}}$ (e.g., \citealt{steffen06, lusso17}). We calculated the IR excess for our observations with this change in assumption and ran the 2D-KS analysis again with this different set of measurements. The 2D-KS probabilities are uniformly lower, typically by $\sim0.2-0.3$\,dex, well within the standard deviations of the probability distributions (e.g., see range of the uniform model in Fig. \ref{sim_stats}). The trends regarding agreement between mock observables and the data remain constant.

 Second, we relaxed the assumption of a linear scaling between IR luminosity and covering fraction. We used the IR-to-bolometric luminosity dependence on covering fraction of \citeauthor{stalevski16} (\citeyear{stalevski16}; an interpolation of the 60$^{\circ}$ line of their Fig. 10). The relation between IR luminosity, covering fraction, and bolometric luminosity then becomes ${\rm L_{\rm IR} = \eta \times BC(f_{\rm cov})\times L_{\rm bol}}$, where BC($f_{\rm cov}$) is the Stalevski dependence. Our Equation \ref{eqn4} then becomes ${\rm log(IR~Excess) = log( BC[{\it f}_{\rm cov}]\,/<BC[{\it f}_{\rm cov;~Type~1}]>)}$. We ran the 2D-KS analysis using this altered empirical relation using both the original IR excess measurements and the variation described above. Adding this nonlinear dependence results in a minor improvement for many of the models at a level of $\sim0.2-0.4$\,dex, again well within the uncertainty range of the probability distribution. When both variations are put in, the changes in the probability map (i.e., Fig. \ref{sim_stats}) mostly cancel out. As a result, these alterations in the definition of IR excess do not change our conclusions that broad distributions of covering fractions results in mock observables with the best agreement with our data.\\\\\\
 
\bibliography{bib}

\vspace{0.5in}
\renewcommand{\thetable}{A\arabic{table}}
\setcounter{table}{0}

\LongTables
\begin{deluxetable*}{lllllll}
\tabletypesize{\scriptsize}
\tablecaption{IR Parameters\label{irpar}}
\tablewidth{\textwidth}
\centering
\tablehead{
\colhead{}  & \colhead{} & \colhead{} & \multicolumn{2}{c}{IR Parameters} \\
\cline{4-5}\\
\colhead{Name} & \colhead{R.A.} &  \colhead{Decl.} & 
\colhead{AGN Slope} & \colhead{Turnover} \\ 
\colhead{} & \colhead{(J2000)} &\colhead{(J2000)} & 
\colhead{($\alpha$)} &  \colhead{Wavelength ($\lambda_{\rm C}$)} 
}
\startdata
LEDA136991 & 00$^{\rm h}$25$^{\rm m}$32.87$^{\rm s}$ & +68$^{\rm d}$21$^{\rm m}$44.2$^{\rm s}$ & $~\,~~1.5^{+0.53}_{-0.40}$ & $~\,~~45.40^{+16.3}_{-15.9}$ \\ 
NGC262 & 00$^{\rm h}$48$^{\rm m}$47.14$^{\rm s}$ & +31$^{\rm d}$57$^{\rm m}$25.1$^{\rm s}$ & $~\,~~1.6^{+0.52}_{-0.45}$ & $~\,~~43.16^{+13.0}_{-12.9}$ \\ 
ESO 195-IG021 & 01$^{\rm h}$00$^{\rm m}$36.53$^{\rm s}$ & -47$^{\rm d}$52$^{\rm m}$02.7$^{\rm s}$ & $~\,~~1.7^{+0.40}_{-0.32}$ & $~\,~~57.95^{+16.8}_{-15.5}$ \\ 
IC 1663 & 01$^{\rm h}$14$^{\rm m}$07.02$^{\rm s}$ & -32$^{\rm d}$39$^{\rm m}$03.2$^{\rm s}$ & $~\,~~2.9^{+0.30}_{-0.21}$ & $~\,~~67.68^{+8.65}_{-9.27}$ \\ 
NGC513 & 01$^{\rm h}$24$^{\rm m}$26.85$^{\rm s}$ & +33$^{\rm d}$47$^{\rm m}$58.0$^{\rm s}$ & $~\,~~1.1^{+0.54}_{-0.43}$ & $~\,~~43.52^{+19.7}_{-15.4}$ \\ 
MCG-01-05-047 & 01$^{\rm h}$52$^{\rm m}$49.00$^{\rm s}$ & -03$^{\rm d}$26$^{\rm m}$48.6$^{\rm s}$ & $~\,~~2.0^{+0.25}_{-0.23}$ & $~\,~~74.65^{+12.1}_{-11.6}$ \\ 
NGC788 & 02$^{\rm h}$01$^{\rm m}$06.45$^{\rm s}$ & -06$^{\rm d}$48$^{\rm m}$55.9$^{\rm s}$ & $~\,~~1.7^{+0.38}_{-0.30}$ & $~\,~~52.66^{+8.73}_{-7.20}$ \\ 
NGC1052 & 02$^{\rm h}$41$^{\rm m}$04.80$^{\rm s}$ & -08$^{\rm d}$15$^{\rm m}$20.7$^{\rm s}$ & $~\,~~1.6^{+0.35}_{-0.27}$ & $~\,~~58.77^{+12.1}_{-9.00}$ \\ 
2MFGC 2280 & 02$^{\rm h}$50$^{\rm m}$42.59$^{\rm s}$ & +54$^{\rm d}$42$^{\rm m}$17.6$^{\rm s}$ & $~\,~~1.7^{+0.64}_{-0.47}$ & $~\,~~43.73^{+23.3}_{-16.7}$ \\ 
NGC1365 & 03$^{\rm h}$33$^{\rm m}$36.37$^{\rm s}$ & -36$^{\rm d}$08$^{\rm m}$25.4$^{\rm s}$ & $~\,~~2.1^{+0.40}_{-0.41}$ & $~\,~~54.32^{+17.6}_{-14.8}$ \\ 
2MASXJ04234080+0408017 & 04$^{\rm h}$23$^{\rm m}$40.77$^{\rm s}$ & +04$^{\rm d}$08$^{\rm m}$01.8$^{\rm s}$ & $~\,~~1.7^{+0.55}_{-0.30}$ & $~\,~~50.48^{+28.2}_{-12.3}$ \\ 
CGCG420-015 & 04$^{\rm h}$53$^{\rm m}$25.75$^{\rm s}$ & +04$^{\rm d}$03$^{\rm m}$41.7$^{\rm s}$ & $~\,~~1.7^{+0.51}_{-0.51}$ & $~\,~~41.75^{+11.1}_{-7.76}$ \\ 
ESO 033-G002 & 04$^{\rm h}$55$^{\rm m}$58.96$^{\rm s}$ & -75$^{\rm d}$32$^{\rm m}$28.2$^{\rm s}$ & $~\,~~1.5^{+0.40}_{-0.33}$ & $~\,~~52.95^{+14.2}_{-11.4}$ \\ 
LEDA178130 & 05$^{\rm h}$05$^{\rm m}$45.73$^{\rm s}$ & -23$^{\rm d}$51$^{\rm m}$14.0$^{\rm s}$ & $~\,~~1.6^{+0.46}_{-0.35}$ & $~\,~~52.21^{+13.7}_{-9.95}$ \\ 
2MASXJ05081967+1721483 & 05$^{\rm h}$08$^{\rm m}$19.69$^{\rm s}$ & +17$^{\rm d}$21$^{\rm m}$48.1$^{\rm s}$ & $~\,~~2.1^{+0.49}_{-0.39}$ & $~\,~~50.01^{+20.3}_{-18.2}$ \\ 
NGC2110 & 05$^{\rm h}$52$^{\rm m}$11.38$^{\rm s}$ & -07$^{\rm d}$27$^{\rm m}$22.3$^{\rm s}$ & $~\,~~1.4^{+0.56}_{-0.42}$ & $~\,~~44.91^{+21.2}_{-17.3}$ \\ 
ESO 005-G004 & 06$^{\rm h}$05$^{\rm m}$41.63$^{\rm s}$ & -86$^{\rm d}$37$^{\rm m}$54.7$^{\rm s}$ & $~\,~~1.0^{+0.48}_{-0.38}$ & $~\,~~51.55^{+18.5}_{-19.0}$ \\ 
ESO 121-IG028 & 06$^{\rm h}$23$^{\rm m}$45.57$^{\rm s}$ & -60$^{\rm d}$58$^{\rm m}$44.4$^{\rm s}$ & $~\,~~1.7^{+0.64}_{-0.48}$ & $~\,~~39.81^{+15.6}_{-13.1}$ \\ 
MCG+06-16-028 & 07$^{\rm h}$14$^{\rm m}$03.86$^{\rm s}$ & +35$^{\rm d}$16$^{\rm m}$45.4$^{\rm s}$ & $~\,~~1.4^{+0.54}_{-0.41}$ & $~\,~~42.86^{+19.1}_{-13.6}$ \\ 
LEDA96373 & 07$^{\rm h}$26$^{\rm m}$26.35$^{\rm s}$ & -35$^{\rm d}$54$^{\rm m}$21.7$^{\rm s}$ & $~\,~~2.0^{+0.46}_{-0.39}$ & $~\,~~49.08^{+11.2}_{-7.22}$ \\ 
UGC3995A & 07$^{\rm h}$44$^{\rm m}$06.97$^{\rm s}$ & +29$^{\rm d}$14$^{\rm m}$56.9$^{\rm s}$ & $~\,~~1.1^{+0.52}_{-0.42}$ & $~\,~~48.55^{+20.2}_{-17.9}$ \\ 
Mrk 1210 & 08$^{\rm h}$04$^{\rm m}$05.86$^{\rm s}$ & +05$^{\rm d}$06$^{\rm m}$49.8$^{\rm s}$ & $~\,~~3.2^{+0.60}_{-0.61}$ & $~\,~~30.46^{+4.67}_{-4.67}$ \\ 
MCG-01-22-006 & 08$^{\rm h}$23$^{\rm m}$01.10$^{\rm s}$ & -04$^{\rm d}$56$^{\rm m}$05.5$^{\rm s}$ & $~\,~~0.85^{+0.45}_{-0.40}$ & $~\,~~49.38^{+18.8}_{-16.8}$ \\ 
MCG+11-11-032 & 08$^{\rm h}$55$^{\rm m}$12.54$^{\rm s}$ & +64$^{\rm d}$23$^{\rm m}$45.6$^{\rm s}$ & $~\,~~1.3^{+0.29}_{-0.27}$ & $~\,~~63.38^{+13.1}_{-10.2}$ \\ 
Mrk 18 & 09$^{\rm h}$01$^{\rm m}$58.39$^{\rm s}$ & +60$^{\rm d}$09$^{\rm m}$06.2$^{\rm s}$ & $~\,~~2.4^{+0.20}_{-0.40}$ & $~\,~~78.11^{+8.24}_{-32.2}$ \\ 
IC 2461 & 09$^{\rm h}$19$^{\rm m}$58.03$^{\rm s}$ & +37$^{\rm d}$11$^{\rm m}$28.5$^{\rm s}$ & $~\,~~2.3^{+0.29}_{-0.47}$ & $~\,~~70.80^{+10.8}_{-9.63}$ \\ 
MCG-01-24-012 & 09$^{\rm h}$20$^{\rm m}$46.25$^{\rm s}$ & -08$^{\rm d}$03$^{\rm m}$22.1$^{\rm s}$ & $~\,~~2.3^{+0.49}_{-0.46}$ & $~\,~~43.76^{+9.25}_{-6.00}$ \\ 
2MASXJ09235371-3141305 & 09$^{\rm h}$23$^{\rm m}$53.73$^{\rm s}$ & -31$^{\rm d}$41$^{\rm m}$30.7$^{\rm s}$ & $~\,~~1.6^{+0.48}_{-0.43}$ & $~\,~~46.95^{+19.8}_{-16.7}$ \\ 
NGC2992 & 09$^{\rm h}$45$^{\rm m}$42.05$^{\rm s}$ & -14$^{\rm d}$19$^{\rm m}$34.9$^{\rm s}$ & $~\,~~2.0^{+0.49}_{-0.39}$ & $~\,~~47.76^{+23.0}_{-18.2}$ \\ 
NGC3079 & 10$^{\rm h}$01$^{\rm m}$57.80$^{\rm s}$ & +55$^{\rm d}$40$^{\rm m}$47.2$^{\rm s}$ & $~\,~~1.1^{+0.50}_{-0.41}$ & $~\,~~51.30^{+18.1}_{-18.0}$ \\ 
ESO 263-G013 & 10$^{\rm h}$09$^{\rm m}$48.21$^{\rm s}$ & -42$^{\rm d}$48$^{\rm m}$40.4$^{\rm s}$ & $~\,~~1.7^{+0.45}_{-0.40}$ & $~\,~~49.35^{+11.8}_{-7.84}$ \\ 
NGC3281 & 10$^{\rm h}$31$^{\rm m}$52.09$^{\rm s}$ & -34$^{\rm d}$51$^{\rm m}$13.3$^{\rm s}$ & $~\,~~2.1^{+0.65}_{-0.59}$ & $~\,~~34.51^{+12.6}_{-9.50}$ \\ 
MCG+12-10-067 & 10$^{\rm h}$44$^{\rm m}$08.54$^{\rm s}$ & +70$^{\rm d}$24$^{\rm m}$19.3$^{\rm s}$ & $~\,~~1.8^{+0.35}_{-0.26}$ & $~\,~~59.41^{+14.0}_{-10.7}$ \\ 
MCG+06-24-008 & 10$^{\rm h}$44$^{\rm m}$48.97$^{\rm s}$ & +38$^{\rm d}$10$^{\rm m}$51.6$^{\rm s}$ & $~\,~~1.1^{+0.57}_{-0.45}$ & $~\,~~41.94^{+19.8}_{-14.9}$ \\ 
UGC5881 & 10$^{\rm h}$46$^{\rm m}$42.52$^{\rm s}$ & +25$^{\rm d}$55$^{\rm m}$53.6$^{\rm s}$ & $~\,~~2.3^{+0.28}_{-0.30}$ & $~\,~~60.28^{+12.5}_{-12.7}$ \\ 
NGC3393 & 10$^{\rm h}$48$^{\rm m}$23.46$^{\rm s}$ & -25$^{\rm d}$09$^{\rm m}$43.4$^{\rm s}$ & $~\,~~2.1^{+0.44}_{-0.38}$ & $~\,~~50.11^{+12.0}_{-7.45}$ \\ 
Mrk 728 & 11$^{\rm h}$01$^{\rm m}$01.78$^{\rm s}$ & +11$^{\rm d}$02$^{\rm m}$48.9$^{\rm s}$ & $~\,~~1.7^{+0.36}_{-0.29}$ & $~\,~~59.45^{+14.5}_{-11.5}$ \\ 
2MASXJ11364205-6003070 & 11$^{\rm h}$36$^{\rm m}$42.05$^{\rm s}$ & -60$^{\rm d}$03$^{\rm m}$06.7$^{\rm s}$ & $~\,~~1.9^{+0.47}_{-0.40}$ & $~\,~~47.47^{+17.7}_{-15.6}$ \\ 
NGC3786 & 11$^{\rm h}$39$^{\rm m}$42.55$^{\rm s}$ & +31$^{\rm d}$54$^{\rm m}$33.4$^{\rm s}$ & $~\,~~1.1^{+0.48}_{-0.43}$ & $~\,~~47.98^{+18.0}_{-16.6}$ \\ 
NGC4388 & 12$^{\rm h}$25$^{\rm m}$46.75$^{\rm s}$ & +12$^{\rm d}$39$^{\rm m}$43.5$^{\rm s}$ & $~\,~~2.1^{+0.58}_{-0.46}$ & $~\,~~43.44^{+15.2}_{-12.1}$ \\ 
LEDA170194 & 12$^{\rm h}$39$^{\rm m}$06.28$^{\rm s}$ & -16$^{\rm d}$10$^{\rm m}$47.1$^{\rm s}$ & $~\,~~1.6^{+0.37}_{-0.31}$ & $~\,~~56.67^{+15.8}_{-13.4}$ \\ 
NGC4941 & 13$^{\rm h}$04$^{\rm m}$13.14$^{\rm s}$ & -05$^{\rm d}$33$^{\rm m}$05.8$^{\rm s}$ & $~\,~~1.5^{+0.29}_{-0.24}$ & $~\,~~65.13^{+12.7}_{-9.78}$ \\ 
NGC4992 & 13$^{\rm h}$09$^{\rm m}$05.60$^{\rm s}$ & +11$^{\rm d}$38$^{\rm m}$03.0$^{\rm s}$ & $~\,~~1.3^{+0.40}_{-0.31}$ & $~\,~~54.30^{+12.4}_{-9.12}$ \\ 
Mrk 248 & 13$^{\rm h}$15$^{\rm m}$17.27$^{\rm s}$ & +44$^{\rm d}$24$^{\rm m}$25.6$^{\rm s}$ & $~\,~~2.0^{+0.71}_{-0.46}$ & $~\,~~41.76^{+32.7}_{-14.7}$ \\ 
ESO 509-IG066 & 13$^{\rm h}$34$^{\rm m}$40.40$^{\rm s}$ & -23$^{\rm d}$26$^{\rm m}$46.0$^{\rm s}$ & $~\,~~2.8^{+0.65}_{-0.64}$ & $~\,~~32.25^{+9.56}_{-7.53}$ \\ 
NGC5252 & 13$^{\rm h}$38$^{\rm m}$15.96$^{\rm s}$ & +04$^{\rm d}$32$^{\rm m}$33.3$^{\rm s}$ & $~\,~~0.91^{+0.41}_{-0.41}$ & $~\,~~55.65^{+17.8}_{-18.5}$ \\ 
NGC5273 & 13$^{\rm h}$42$^{\rm m}$08.34$^{\rm s}$ & +35$^{\rm d}$39$^{\rm m}$15.2$^{\rm s}$ & $~\,~~1.3^{+0.47}_{-0.43}$ & $~\,~~53.76^{+17.1}_{-18.4}$ \\ 
NGC5674 & 14$^{\rm h}$33$^{\rm m}$52.24$^{\rm s}$ & +05$^{\rm d}$27$^{\rm m}$29.6$^{\rm s}$ & $~\,~~0.44^{+0.53}_{-0.39}$ & $~\,~~48.19^{+19.6}_{-18.9}$ \\ 
NGC5728 & 14$^{\rm h}$42$^{\rm m}$23.90$^{\rm s}$ & -17$^{\rm d}$15$^{\rm m}$11.1$^{\rm s}$ & $~\,~~2.0^{+0.39}_{-0.45}$ & $~\,~~63.18^{+11.5}_{-11.3}$ \\ 
IC 4518A & 14$^{\rm h}$57$^{\rm m}$41.18$^{\rm s}$ & -43$^{\rm d}$07$^{\rm m}$55.6$^{\rm s}$ & $~\,~~2.6^{+0.59}_{-0.43}$ & $~\,~~45.09^{+17.2}_{-15.6}$ \\ 
2MASXJ15064412+0351444 & 15$^{\rm h}$06$^{\rm m}$44.13$^{\rm s}$ & +03$^{\rm d}$51$^{\rm m}$44.4$^{\rm s}$ & $~\,~~1.6^{+0.50}_{-0.50}$ & $~\,~~51.64^{+18.1}_{-17.2}$ \\ 
NGC5899 & 15$^{\rm h}$15$^{\rm m}$03.22$^{\rm s}$ & +42$^{\rm d}$02$^{\rm m}$59.5$^{\rm s}$ & $~\,~~1.2^{+0.47}_{-0.44}$ & $~\,~~63.82^{+14.9}_{-17.4}$ \\ 
MCG+11-19-006 & 15$^{\rm h}$19$^{\rm m}$33.69$^{\rm s}$ & +65$^{\rm d}$35$^{\rm m}$58.5$^{\rm s}$ & $~\,~~1.8^{+0.76}_{-0.56}$ & $~\,~~34.46^{+14.6}_{-12.0}$ \\ 
MCG-01-40-001 & 15$^{\rm h}$33$^{\rm m}$20.71$^{\rm s}$ & -08$^{\rm d}$42$^{\rm m}$01.9$^{\rm s}$ & $~\,~~2.8^{+0.42}_{-0.29}$ & $~\,~~54.32^{+11.6}_{-8.39}$ \\ 
NGC5995 & 15$^{\rm h}$48$^{\rm m}$24.96$^{\rm s}$ & -13$^{\rm d}$45$^{\rm m}$27.9$^{\rm s}$ & $~\,~~1.6^{+0.45}_{-0.34}$ & $~\,~~50.36^{+16.3}_{-12.6}$ \\ 
MCG+14-08-004 & 16$^{\rm h}$19$^{\rm m}$19.26$^{\rm s}$ & +81$^{\rm d}$02$^{\rm m}$48.6$^{\rm s}$ & $~\,~~1.6^{+0.55}_{-0.45}$ & $~\,~~42.77^{+12.1}_{-13.1}$ \\ 
NGC6240 & 16$^{\rm h}$52$^{\rm m}$58.87$^{\rm s}$ & +02$^{\rm d}$24$^{\rm m}$03.3$^{\rm s}$ & $~\,~~2.8^{+0.87}_{-0.44}$ & $~\,~~40.29^{+34.8}_{-12.9}$ \\ 
NGC6300 & 17$^{\rm h}$16$^{\rm m}$59.47$^{\rm s}$ & -62$^{\rm d}$49$^{\rm m}$14.0$^{\rm s}$ & $~\,~~1.7^{+0.30}_{-0.32}$ & $~\,~~65.46^{+12.6}_{-10.3}$ \\ 
MCG+07-37-031 & 18$^{\rm h}$16$^{\rm m}$11.55$^{\rm s}$ & +42$^{\rm d}$39$^{\rm m}$37.2$^{\rm s}$ & $~\,~~2.3^{+0.51}_{-0.35}$ & $~\,~~47.79^{+21.5}_{-14.8}$ \\ 
IC 4709 & 18$^{\rm h}$24$^{\rm m}$19.39$^{\rm s}$ & -56$^{\rm d}$22$^{\rm m}$09.0$^{\rm s}$ & $~\,~~1.8^{+0.40}_{-0.32}$ & $~\,~~52.87^{+11.4}_{-8.34}$ \\ 
ESO 103-G035 & 18$^{\rm h}$38$^{\rm m}$20.34$^{\rm s}$ & -65$^{\rm d}$25$^{\rm m}$39.2$^{\rm s}$ & $~\,~~2.9^{+0.69}_{-0.69}$ & $~\,~~29.96^{+5.93}_{-5.67}$ \\ 
2MASXJ20183871+4041003 & 20$^{\rm h}$18$^{\rm m}$38.72$^{\rm s}$ & +40$^{\rm d}$41$^{\rm m}$00.2$^{\rm s}$ & $~\,~~0.93^{+0.56}_{-0.42}$ & $~\,~~44.93^{+19.1}_{-17.3}$ \\ 
MCG+04-48-002 & 20$^{\rm h}$28$^{\rm m}$35.06$^{\rm s}$ & +25$^{\rm d}$44$^{\rm m}$00.0$^{\rm s}$ & $~\,~~1.3^{+0.53}_{-0.43}$ & $~\,~~47.11^{+18.7}_{-17.9}$ \\ 
IC 5063 & 20$^{\rm h}$52$^{\rm m}$02.34$^{\rm s}$ & -57$^{\rm d}$04$^{\rm m}$07.6$^{\rm s}$ & $~\,~~2.2^{+0.47}_{-0.48}$ & $~\,~~43.60^{+9.12}_{-5.44}$ \\ 
MCG+06-49-019 & 22$^{\rm h}$27$^{\rm m}$05.78$^{\rm s}$ & +36$^{\rm d}$21$^{\rm m}$41.7$^{\rm s}$ & $~\,~~1.5^{+0.34}_{-0.29}$ & $~\,~~60.78^{+13.1}_{-10.8}$ \\ 
MCG+01-57-016 & 22$^{\rm h}$40$^{\rm m}$17.05$^{\rm s}$ & +08$^{\rm d}$03$^{\rm m}$14.1$^{\rm s}$ & $~\,~~1.9^{+0.37}_{-0.33}$ & $~\,~~54.71^{+13.4}_{-12.2}$ \\ 
NGC7582 & 23$^{\rm h}$18$^{\rm m}$23.50$^{\rm s}$ & -42$^{\rm d}$22$^{\rm m}$14.0$^{\rm s}$ & $~\,~~2.1^{+0.59}_{-0.43}$ & $~\,~~43.57^{+20.5}_{-14.9}$ \\ 
2MASXJ23303771+7122464 & 23$^{\rm h}$30$^{\rm m}$37.69$^{\rm s}$ & +71$^{\rm d}$22$^{\rm m}$46.5$^{\rm s}$ & $~\,~~1.6^{+0.63}_{-0.45}$ & $~\,~~42.37^{+19.9}_{-16.9}$ \\ 
PKS 2331-240 & 23$^{\rm h}$33$^{\rm m}$55.24$^{\rm s}$ & -23$^{\rm d}$43$^{\rm m}$40.66$^{\rm s}$ & $~\,~~1.2^{+0.12}_{-0.11}$ & $~\,~~137.7^{+9.34}_{-10.5}$ 
\enddata
\tablecomments{Names (Column 1) and coordinates (Columns 2 and 3) of our sample, along with two parameters from the SED decomposition from S17: the slope of the exponentially cutoff power law (Column 4) and its turnover wavelength (Column 5). Further details of the modeling are given in Section 3.2.}
\end{deluxetable*}

\LongTables
\begin{deluxetable*}{lrrlllll}
\tabletypesize{\scriptsize}
\tablecaption{NuSTAR Parameters\label{nustar}}
\tablewidth{\textwidth}
 \centering
\tablehead{
\colhead{}  & \multicolumn{5}{c}{NuSTAR parameters} \\
\cline{2-6}\\
\colhead{Name} & \colhead{Log(N$_{\rm H}$)} & \colhead{Gamma} & 
\colhead{EW(Fe\,K$\alpha$)} & \colhead{Reflection} & \colhead{Unabs. PL} \\
\colhead{} & \colhead{(cm$^{-2}$)} &  \colhead{} &
\colhead{(keV)}  & \colhead{Parameter} & \colhead{Normalization}  }
\startdata
LEDA136991 & $~\,~~23.95^{+0.17}_{-0.18}$ & $1.711^{+0.138}_{-0.127}$ & $~~~(7.50^{+6.00}_{-1.96})\times10^{-1}$ & $~~~~~2.21^{+2.69}_{-1.11}$ & $<9.80\times10^{-1}$ \\ 
NGC262 & $~\,~~22.97^{+0.01}_{-0.01}$ & $1.747^{+0.020}_{-0.019}$ & $~~~(4.47^{+1.17}_{-1.18})\times10^{-2}$ & $~~~(5.79^{+0.78}_{-0.75})\times10^{-1}$ & $~~~(2.06^{+0.92}_{-0.91})\times10^{-1}$ \\ 
ESO 195-IG021 & $~\,~~22.62^{+0.04}_{-0.05}$ & $1.876^{+0.078}_{-0.073}$ & $~~~(1.35^{+0.35}_{-0.36})\times10^{-1}$ & $~~~(5.15^{+3.63}_{-2.86})\times10^{-1}$ & $~~~(4.83^{+3.71}_{-3.65})\times10^{-1}$ \\ 
IC 1663 & $~\,~~23.44^{+0.09}_{-0.11}$ & $1.571^{+0.245}_{-0.168}$ & $~~~(4.04^{+14.6}_{l})\times10^{-2}$ & $~~~(1.72^{+7.41}_{l})\times10^{-1}$ & $~~~~~1.62^{+2.18}_{-1.00}$ \\ 
NGC513 & $~\,~~22.85^{+0.08}_{-0.08}$ & $1.699^{+0.117}_{-0.111}$ & $~~~(1.75^{+0.62}_{-0.63})\times10^{-1}$ & $~~~~~1.03^{+0.64}_{-0.46}$ & $~~~~~2.49^{+1.16}_{-0.92}$ \\ 
MCG-01-05-047 & $~\,~~23.22^{+0.06}_{-0.07}$ & $1.807^{+0.097}_{-0.101}$ & $~~~(3.38^{+0.88}_{-0.84})\times10^{-1}$ & $~~~~~2.86^{+1.27}_{-0.90}$ & $~~~~~1.32^{+0.68}_{-0.56}$ \\ 
NGC788 & $~\,~~23.89^{+0.04}_{-0.03}$ & $1.770^{+0.047}_{-0.046}$ & $~~~(2.92^{+0.77}_{-0.72})\times10^{-1}$ & $~~~(7.45^{+2.34}_{-1.84})\times10^{-1}$ & $~~~~~1.25^{+0.46}_{-0.36}$ \\ 
NGC1052 & $~\,~~23.04^{+0.04}_{-0.05}$ & $1.516^{+0.042}_{-0.040}$ & $~~~(1.63^{+0.43}_{-0.43})\times10^{-1}$ & $<3.30\times10^{-1}$ & $~~~~~8.12^{+1.53}_{-1.29}$ \\ 
2MFGC 2280 & $~\,~~24.20^{+0.07}_{-0.06}$ & $1.564^{+0.161}_{-0.154}$ & $~~~(1.28^{+1.03}_{-0.84})\times10^{0}$ & $~~~(5.78^{+9.12}_{l})\times10^{-2}$ & $~~~(2.21^{+2.34}_{-1.32})\times10^{-1}$ \\ 
NGC1365 & $~\,~~23.30^{+0.02}_{-0.02}$ & $1.903^{+0.034}_{-0.033}$ & $~~~(8.08^{+1.87}_{-1.88})\times10^{-2}$ & $~~~~~2.98^{+0.41}_{-0.36}$ & $~~~~~4.13^{+1.06}_{-0.96}$ \\ 
2MASXJ04234080+0408017 & $~\,~~23.90^{+0.04}_{-0.04}$ & $1.769^{+0.083}_{-0.092}$ & $~~~(1.65^{+7.43}_{l})\times10^{-2}$ & $~~~(3.95^{+1.87}_{-1.70})\times10^{-1}$ & $~~~~~1.11^{+0.57}_{-0.38}$ \\ 
CGCG420-015 & $~\,~~23.98^{+0.04}_{-0.05}$ & $1.885^{+0.057}_{-0.058}$ & $~~~(4.92^{+0.83}_{-0.66})\times10^{-1}$ & $~~~~~1.14^{+0.37}_{-0.27}$ & $~~~(3.72^{+5.05}_{l})\times10^{-1}$ \\ 
ESO 033-G002 & $~\,~~22.26^{+0.03}_{-0.03}$ & $2.173^{+0.067}_{-0.061}$ & $~~~(9.41^{+2.39}_{-2.41})\times10^{-2}$ & $~~~~~2.34^{+0.64}_{-0.51}$ & $~~~(4.77^{+3.33}_{-3.19})\times10^{-1}$ \\ 
LEDA178130 & $~\,~~23.05^{+0.03}_{-0.02}$ & $1.667^{+0.047}_{-0.023}$ & $~~~(6.41^{+2.07}_{-2.11})\times10^{-2}$ & $~~~(2.75^{+12.0}_{l})\times10^{-2}$ & $~~~(7.41^{+2.21}_{-2.07})\times10^{-1}$ \\ 
2MASXJ05081967+1721483 & $~\,~~22.21^{+0.04}_{-0.04}$ & $1.738^{+0.062}_{-0.059}$ & $~~~(1.44^{+0.31}_{-0.31})\times10^{-1}$ & $~~~(4.90^{+3.00}_{-2.44})\times10^{-1}$ & $~~~~~1.19^{+0.88}_{-0.89}$ \\ 
NGC2110 & $~\,~~22.58^{+0.01}_{-0.01}$ & $1.640^{+0.010}_{-0.008}$ & $~~~(3.26^{+0.59}_{-0.59})\times10^{-2}$ & $~~~(1.73^{+2.79}_{l})\times10^{-2}$ & $~~~(4.77^{+1.61}_{-1.61})\times10^{-1}$ \\ 
ESO 005-G004 & $~\,~~24.23^{+0.30}_{-0.14}$ & $1.8^{f}$ & $~~~(1.67^{+1.50}_{-0.41})\times10^{0}$ & $~~~~~1.88^{+2.39}_{-1.32}$ & $~~~~~1.07^{+1.19}_{-0.79}$ \\ 
ESO 121-IG028 & $~\,~~23.36^{+0.04}_{-0.04}$ & $1.832^{+0.084}_{-0.086}$ & $~~~(5.98^{+4.12}_{-4.14})\times10^{-2}$ & $~~~(6.93^{+3.37}_{-2.78})\times10^{-1}$ & $<2.10\times10^{-1}$ \\ 
MCG+06-16-028 & $~\,~~24.15^{+0.08}_{-0.06}$ & $1.792^{+0.157}_{-0.104}$ & $~~~(4.02^{+1.90}_{-1.43})\times10^{-1}$ & $~~~(3.62^{+2.01}_{-1.78})\times10^{-1}$ & $~~~(9.44^{+6.46}_{-4.81})\times10^{-1}$ \\ 
LEDA96373 & $~\,~~24.10^{f}$ & $1.957^{+0.070}_{-0.078}$ & $~~~(8.05^{+4.35}_{-1.28})\times10^{-1}$ & $~~~~~2.11^{+2.13}_{-0.93}$ & $~~~~~3.39^{+2.45}_{-1.48}$ \\ 
UGC3995A & $~\,~~23.59^{+0.06}_{-0.05}$ & $1.737^{+0.075}_{-0.080}$ & $~~~(1.48^{+0.54}_{-0.52})\times10^{-1}$ & $~~~~~2.01^{+0.71}_{-0.54}$ & $~~~(6.68^{+5.12}_{-4.02})\times10^{-1}$ \\ 
Mrk 1210 & $~\,~~23.43^{+0.02}_{-0.03}$ & $1.876^{+0.050}_{-0.052}$ & $~~~(1.13^{+0.29}_{-0.30})\times10^{-1}$ & $~~~~~1.65^{+0.34}_{-0.31}$ & $~~~~~1.71^{+0.41}_{-0.35}$ \\ 
MCG-01-22-006 & $~\,~~23.30^{+0.02}_{-0.03}$ & $1.560^{+0.064}_{-0.061}$ & $~~~(6.29^{+2.84}_{-2.84})\times10^{-2}$ & $~~~(4.44^{+2.06}_{-1.73})\times10^{-1}$ & $~~~(5.85^{+3.37}_{-2.97})\times10^{-1}$ \\ 
MCG+11-11-032 & $~\,~~23.07^{+0.09}_{-0.09}$ & $1.866^{+0.167}_{-0.160}$ & $~~~(3.90^{+9.00}_{l})\times10^{-2}$ & $~~~~~1.40^{+1.27}_{-0.77}$ & $<1.25\times10^{0}$ \\ 
Mrk 18 & $~\,~~23.11^{+0.10}_{-0.13}$ & $1.627^{+0.201}_{-0.114}$ & $~~~(1.76^{+1.14}_{-1.10})\times10^{-1}$ & $~~~(1.03^{+5.61}_{l})\times10^{-1}$ & $~~~~~2.32^{+2.65}_{-1.30}$ \\ 
IC 2461 & $~\,~~22.86^{+0.06}_{-0.06}$ & $1.802^{+0.097}_{-0.093}$ & $~~~(1.18^{+0.40}_{-0.39})\times10^{-1}$ & $~~~(6.95^{+4.35}_{-3.38})\times10^{-1}$ & $<3.50\times10^{-1}$ \\ 
MCG-01-24-012 & $~\,~~22.97^{+0.02}_{-0.03}$ & $2.074^{+0.061}_{-0.060}$ & $~~~(4.03^{+2.15}_{-2.15})\times10^{-2}$ & $~~~~~1.29^{+0.38}_{-0.31}$ & $~~~(5.78^{+15.2}_{l})\times10^{-2}$ \\ 
2MASXJ09235371-3141305 & $~\,~~23.89^{+0.08}_{-0.09}$ & $1.866^{+0.163}_{-0.176}$ & $~~~(5.54^{+8.06}_{l})\times10^{-2}$ & $~~~(8.12^{+3.88}_{-2.61})\times10^{-1}$ & $~~~(2.03^{+5.45}_{l})\times10^{-1}$ \\ 
NGC2992 & $~\,~~22.04^{+0.02}_{-0.01}$ & $1.724^{+0.018}_{-0.018}$ & $~~~(7.96^{+1.07}_{-1.06})\times10^{-2}$ & $~~~(1.31^{+0.69}_{-0.66})\times10^{-1}$ & $~~~(6.84^{+2.73}_{-2.73})\times10^{-1}$ \\ 
NGC3079 & $~\,~~24.52^{+0.04}_{-0.04}$ & $2.017^{+0.115}_{-0.114}$ & $~~~(3.83^{+2.82}_{-1.92})\times10^{-1}$ & $~~~(2.09^{+0.90}_{-0.62})\times10^{-2}$ & $~~~(7.23^{+6.47}_{-3.70})\times10^{-2}$ \\ 
ESO 263-G013 & $~\,~~23.87^{+0.03}_{-0.04}$ & $1.732^{+0.085}_{-0.085}$ & $~~~(6.24^{+6.46}_{l})\times10^{-2}$ & $~~~(8.42^{+15.2}_{l})\times10^{-2}$ & $~~~~~1.30^{+0.67}_{-0.58}$ \\ 
NGC3281 & $~\,~~24.08^{+0.09}_{-0.10}$ & $1.622^{+0.033}_{-0.032}$ & $~~~(1.09^{+0.79}_{-0.13})\times10^{0}$ & $~~~~~3.72^{+4.38}_{-1.64}$ & $<1.20\times10^{-1}$ \\ 
MCG+12-10-067 & $~\,~~23.24^{+0.07}_{-0.07}$ & $1.923^{+0.155}_{-0.147}$ & $~~~(8.10^{+6.00}_{-6.00})\times10^{-2}$ & $~~~~~1.37^{+0.99}_{-0.65}$ & $~~~~~1.75^{+1.09}_{-0.79}$ \\ 
MCG+06-24-008 & $~\,~~22.60^{+0.08}_{-0.10}$ & $1.564^{+0.047}_{-0.046}$ & $~~~(8.68^{+5.32}_{-5.28})\times10^{-2}$ & $<8.50\times10^{-2}$ & $~~~~~1.09^{+0.87}_{-0.94}$ \\ 
UGC5881 & $~\,~~23.01^{+0.10}_{-0.11}$ & $1.628^{+0.163}_{-0.149}$ & $~~~(1.37^{+0.81}_{-0.80})\times10^{-1}$ & $~~~(7.08^{+8.02}_{-5.04})\times10^{-1}$ & $~~~~~3.97^{+2.78}_{-1.84}$ \\ 
NGC3393 & $~\,~~24.38^{+0.04}_{-0.05}$ & $1.850^{+0.140}_{-0.134}$ & $~~~(6.89^{+4.01}_{-3.37})\times10^{-1}$ & $~~~(3.42^{+2.41}_{-2.16})\times10^{-2}$ & $~~~(2.14^{+1.94}_{-1.19})\times10^{-1}$ \\ 
Mrk 728 & $~\,~~21.86^{+0.40}_{l}$ & $1.591^{+0.055}_{-0.050}$ & $~~~(8.48^{+4.52}_{-4.55})\times10^{-2}$ & $<2.00\times10^{-1}$ &  ...  \\ 
2MASXJ11364205-6003070 & $~\,~~20.59^{+0.48}_{-0.59}$ & $1.996^{+0.085}_{-0.074}$ & $~~~(1.13^{+0.41}_{-0.42})\times10^{-1}$ & $~~~~~1.49^{+0.67}_{-0.50}$ &  ...  \\ 
NGC3786 & $~\,~~22.52^{+0.23}_{-0.42}$ & $1.754^{+0.203}_{-0.185}$ & $~~~(1.41^{+1.10}_{-1.08})\times10^{-1}$ & $~~~(7.27^{+11.5}_{-6.81})\times10^{-1}$ & $~~~~~6.94^{+7.26}_{l}$ \\ 
NGC4388 & $~\,~~23.67^{+0.02}_{-0.02}$ & $1.699^{+0.016}_{-0.015}$ & $~~~(3.00^{+0.30}_{-0.30})\times10^{-1}$ & $~~~(9.05^{+6.15}_{-5.63})\times10^{-2}$ & $~~~~~6.82^{+0.56}_{-0.55}$ \\ 
LEDA170194 & $~\,~~22.75^{+0.07}_{-0.07}$ & $1.719^{+0.090}_{-0.072}$ & $~~~(1.94^{+0.62}_{-0.62})\times10^{-1}$ & $~~~(1.54^{+3.33}_{l})\times10^{-1}$ & $~~~~~3.65^{+1.55}_{-1.24}$ \\ 
NGC4941 & $~\,~~24.14^{+0.07}_{-0.07}$ & $1.738^{+0.157}_{-0.149}$ & $~~~(9.28^{+5.52}_{-3.12})\times10^{-1}$ & $~~~(2.10^{+1.87}_{-1.48})\times10^{-1}$ & $~~~(9.80^{+8.40}_{-5.27})\times10^{-1}$ \\ 
NGC4992 & $~\,~~23.63^{+0.03}_{-0.03}$ & $1.570^{+0.052}_{-0.053}$ & $~~~(1.80^{+0.48}_{-0.48})\times10^{-1}$ & $~~~(6.98^{+2.20}_{-1.84})\times10^{-1}$ & $<1.20\times10^{-1}$ \\ 
Mrk 248 & $~\,~~23.03^{+0.04}_{-0.05}$ & $1.992^{+0.102}_{-0.103}$ & $~~~(6.90^{+3.90}_{-3.86})\times10^{-2}$ & $~~~~~1.19^{+0.61}_{-0.49}$ & $~~~(3.27^{+2.35}_{-2.03})\times10^{-1}$ \\ 
ESO 509-IG066 & $~\,~~22.89^{+0.06}_{-0.07}$ & $1.704^{+0.118}_{-0.114}$ & $~~~(1.81^{+0.46}_{-0.45})\times10^{-1}$ & $~~~(4.89^{+4.81}_{-3.55})\times10^{-1}$ & $~~~~~1.34^{+0.79}_{-0.67}$ \\ 
NGC5252 & $~\,~~22.55^{+0.06}_{-0.07}$ & $1.662^{+0.023}_{-0.023}$ & $~~~(8.17^{+2.53}_{-2.51})\times10^{-2}$ & $<2.50\times10^{-2}$ &  ...  \\ 
NGC5273 & $~\,~~22.43^{+0.03}_{-0.04}$ & $1.797^{+0.049}_{-0.046}$ & $~~~(1.10^{+0.20}_{-0.21})\times10^{-1}$ & $~~~~~1.07^{+0.29}_{-0.24}$ & $~~~~~3.62^{+0.59}_{-0.54}$ \\ 
NGC5674 & $~\,~~22.66^{+0.05}_{-0.04}$ & $1.871^{+0.086}_{-0.078}$ & $~~~(1.48^{+0.36}_{-0.35})\times10^{-1}$ & $~~~(4.61^{+3.82}_{-2.87})\times10^{-1}$ & $~~~~~1.41^{+0.47}_{-0.42}$ \\ 
NGC5728 & $~\,~~24.14^{+0.02}_{-0.02}$ & $1.636^{+0.045}_{-0.044}$ & $~~~(3.92^{+0.69}_{-0.65})\times10^{-1}$ & $~~~(1.75^{+0.39}_{-0.35})\times10^{-1}$ & $~~~(4.73^{+10.1}_{l})\times10^{-2}$ \\ 
IC 4518A & $~\,~~23.23^{+0.06}_{-0.05}$ & $1.996^{+0.091}_{-0.085}$ & $<7.40\times10^{-2}$ & $~~~~~2.90^{+1.20}_{-0.79}$ & $~~~~~1.53^{+0.59}_{-0.56}$ \\ 
2MASXJ15064412+0351444 & $~\,~~22.30^{+0.09}_{-0.08}$ & $1.689^{+0.057}_{-0.057}$ & $<1.10\times10^{-1}$ & $<3.00\times10^{-1}$ & $<2.00\times10^{0}$ \\ 
NGC5899 & $~\,~~22.98^{+0.04}_{-0.04}$ & $1.903^{+0.080}_{-0.077}$ & $~~~(1.34^{+0.36}_{-0.35})\times10^{-1}$ & $~~~~~1.13^{+0.45}_{-0.35}$ & $~~~(2.53^{+2.62}_{-2.33})\times10^{-1}$ \\ 
MCG+11-19-006 & $~\,~~23.25^{+0.08}_{-0.09}$ & $1.576^{+0.150}_{-0.146}$ & $~~~(4.41^{+8.99}_{l})\times10^{-2}$ & $~~~(4.97^{+5.83}_{-3.95})\times10^{-1}$ & $~~~~~1.29^{+1.10}_{-0.75}$ \\ 
MCG-01-40-001 & $~\,~~22.81^{+0.05}_{-0.06}$ & $1.790^{+0.087}_{-0.085}$ & $~~~(1.99^{+0.45}_{-0.44})\times10^{-1}$ & $~~~(9.13^{+4.57}_{-3.64})\times10^{-1}$ & $~~~~~4.45^{+1.23}_{-1.00}$ \\ 
NGC5995 & $~\,~~22.09^{+0.03}_{-0.03}$ & $1.992^{+0.047}_{-0.044}$ & $~~~(1.65^{+0.23}_{-0.24})\times10^{-1}$ & $~~~~~1.16^{+0.31}_{-0.26}$ & $~~~~~2.72^{+0.76}_{-0.74}$ \\ 
MCG+14-08-004 & $~\,~~23.14^{+0.08}_{-0.07}$ & $1.696^{+0.132}_{-0.086}$ & $~~~(1.91^{+0.77}_{-0.77})\times10^{-1}$ & $~~~(1.10^{+3.86}_{l})\times10^{-1}$ & $<6.90\times10^{-1}$ \\ 
NGC6240 & $~\,~~24.10^{+0.02}_{-0.02}$ & $1.705^{+0.047}_{-0.047}$ & $~~~(1.54^{+0.42}_{-0.42})\times10^{-1}$ & $~~~(2.55^{+0.70}_{-0.65})\times10^{-1}$ & $<1.50\times10^{-1}$ \\ 
NGC6300 & $~\,~~23.23^{+0.02}_{-0.02}$ & $1.897^{+0.029}_{-0.030}$ & $~~~(4.35^{+1.81}_{-1.80})\times10^{-2}$ & $~~~~~1.47^{+0.18}_{-0.17}$ & $<4.50\times10^{-2}$ \\ 
MCG+07-37-031 & $~\,~~22.56^{+0.06}_{-0.05}$ & $1.681^{+0.072}_{-0.069}$ & $~~~(7.93^{+3.67}_{-3.71})\times10^{-2}$ & $~~~(3.54^{+2.89}_{-2.33})\times10^{-1}$ & $~~~~~2.59^{+0.82}_{-0.73}$ \\ 
IC 4709 & $~\,~~23.42^{+0.05}_{-0.04}$ & $1.927^{+0.071}_{-0.073}$ & $~~~(1.55^{+0.41}_{-0.41})\times10^{-1}$ & $~~~~~2.06^{+0.59}_{-0.49}$ & $<4.30\times10^{-1}$ \\ 
ESO 103-G035 & $~\,~~23.33^{+0.01}_{-0.01}$ & $1.965^{+0.021}_{-0.022}$ & $~~~(8.25^{+1.37}_{-1.35})\times10^{-2}$ & $~~~~~1.19^{+0.11}_{-0.11}$ & $<1.10\times10^{-2}$ \\ 
2MASXJ20183871+4041003 & $~\,~~23.14^{+0.05}_{-0.04}$ & $1.699^{+0.087}_{-0.087}$ & $~~~(1.53^{+0.41}_{-0.40})\times10^{-1}$ & $~~~(9.81^{+4.19}_{-3.35})\times10^{-1}$ & $<3.00\times10^{0}$ \\ 
MCG+04-48-002 & $~\,~~23.95^{+0.08}_{-0.08}$ & $1.764^{+0.146}_{-0.156}$ & $~~~(3.41^{+1.72}_{-1.27})\times10^{-1}$ & $~~~(5.50^{+3.71}_{-2.80})\times10^{-1}$ & $~~~(4.13^{+10.8}_{l})\times10^{-1}$ \\ 
IC 5063 & $~\,~~23.42^{+0.02}_{-0.03}$ & $1.799^{+0.050}_{-0.050}$ & $~~~(1.19^{+0.26}_{-0.25})\times10^{-1}$ & $~~~(8.18^{+2.02}_{-1.85})\times10^{-1}$ & $~~~(5.18^{+1.81}_{-1.61})\times10^{-1}$ \\ 
MCG+06-49-019 & $<21.00$ & $1.7^{f}$ & $~~~(5.02^{+1.73}_{-1.71})\times10^{-1}$ & $~~~(3.31^{+3.49}_{-3.18})\times10^{-1}$ &  ...  \\ 
MCG+01-57-016 & $<20.10$ & $1.850^{+0.052}_{-0.051}$ & $~~~(8.28^{+4.52}_{-4.50})\times10^{-2}$ & $~~~~~1.26^{+0.41}_{-0.36}$ &  ...  \\ 
NGC7582 & $~\,~~23.45^{+0.04}_{-0.05}$ & $1.764^{+0.038}_{-0.038}$ & $~~~(2.48^{+0.42}_{-0.41})\times10^{-1}$ & $~~~~~5.43^{+1.29}_{-0.99}$ & $~~~~~1.77^{+0.97}_{-0.87}$ \\ 
2MASXJ23303771+7122464 & $~\,~~22.86^{+0.15}_{-0.21}$ & $1.665^{+0.194}_{-0.156}$ & $~~~(1.63^{+8.47}_{l})\times10^{-2}$ & $~~~(2.81^{+7.11}_{l})\times10^{-1}$ & $~~~~~2.51^{+3.78}_{l}$ \\ 
PKS 2331-240 & $~\,~~20.81^{+0.06}_{-0.06}$ & $1.811^{+0.020}_{-0.019}$ & $~~~(9.81^{+2.99}_{-2.96})\times10^{-2}$ & $<1.30\times10^{-2}$ &  ...  
\enddata
\tablecomments{Parameters from the \nustar modeling of B18: column density (Column 2), intrinsic power law slope (Column 3), equivalent width of the Fe\,K$\alpha$ line (Column 4), the absolute value of the \texttt{pexrav} reflection parameter (Column 5), and the normalization of the unabsorbed, exponentially cutoff power law (Column 6). Uncertainties given as $l$ indicate that the lower limit of the uncertainty is poorly constrained, despite the fit returning a best value for the parameter. Further details of the modeling are given in Section 3.1.}
\end{deluxetable*}

\LongTables
\begin{deluxetable*}{llllllll}
\tabletypesize{\scriptsize}
\tablecaption{Luminosities\label{lumin}}
\tablecolumns{7}
\tablewidth{0pt}
\centering
\tablehead{
\colhead{}  & \multicolumn{6}{c}{Log(Luminosities/erg\,s$^{-1}$)}& \colhead{} \\
\cline{2-7}\\
\colhead{Name} & \colhead{2-10 keV} &  \colhead{10-50 keV} & 
\colhead{2-10 keV} & \colhead{10-50 keV} & 
 \colhead{AGN IR} & 
\colhead{Total IR} & \colhead{f$_{\rm AGN~IR}$} \\
\colhead{} & \colhead{Observ.} &\colhead{Observ. } & 
\colhead{Intrinsic } &  \colhead{Intrinsic} &
\colhead{} & \colhead{} & \colhead{} }
\startdata
LEDA136991 & $39.78^{+2.44}_{-0.03}$ & $42.25^{+0.03}_{-0.03}$ & $41.85^{+0.26}_{-0.30}$ & $42.05^{+0.24}_{-0.29}$ & $~\,~~43.06^{+0.07}_{-0.11}$ & $~~\,~43.21^{+0.04}_{-0.05}$ & $~\,~~0.70^{+0.11}_{-0.11}$ \\ 
NGC262 & $41.77^{+0.02}_{-0.01}$ & $43.77^{+0.02}_{-0.01}$ & $43.50^{+0.02}_{-0.02}$ & $43.68^{+0.01}_{-0.01}$ & $~\,~~43.82^{+0.07}_{-0.07}$ & $~~\,~43.94^{+0.04}_{-0.04}$ & $~\,~~0.76^{+0.10}_{-0.10}$ \\ 
ESO 195-IG021 & $41.51^{+0.01}_{-0.02}$ & $43.82^{+0.03}_{-0.01}$ & $43.64^{+0.07}_{-0.07}$ & $43.73^{+0.05}_{-0.05}$ & $~\,~~43.89^{+0.18}_{-0.24}$ & $~~\,~44.38^{+0.02}_{-0.02}$ & $~\,~~0.33^{+0.14}_{-0.15}$ \\ 
IC 1663 & $39.89^{+0.46}_{-0.01}$ & $42.05^{+0.36}_{--0.0}$ & $41.79^{+0.19}_{-0.18}$ & $42.09^{+0.08}_{-0.14}$ & $~\,~~43.61^{+0.05}_{-0.06}$ & $~~\,~43.86^{+0.02}_{-0.02}$ & $~\,~~0.57^{+0.10}_{-0.10}$ \\ 
NGC513 & $40.78^{+0.02}_{-0.02}$ & $42.92^{+0.03}_{-0.01}$ & $42.53^{+0.10}_{-0.11}$ & $42.74^{+0.06}_{-0.08}$ & $<43.23$ & $~~\,~44.23^{+0.01}_{-0.02}$ & $<0.10$ \\ 
MCG-01-05-047 & $40.68^{+0.02}_{-0.02}$ & $42.96^{+0.03}_{-0.02}$ & $42.45^{+0.11}_{-0.11}$ & $42.59^{+0.09}_{-0.09}$ & $~\,~~43.71^{+0.13}_{-0.16}$ & $~~\,~44.24^{+0.01}_{-0.02}$ & $~\,~~0.29^{+0.10}_{-0.10}$ \\ 
NGC788 & $40.48^{+0.01}_{-0.02}$ & $42.95^{+0.01}_{-0.02}$ & $42.84^{+0.06}_{-0.07}$ & $43.00^{+0.05}_{-0.06}$ & $~\,~~43.56^{+0.04}_{-0.04}$ & $~~\,~43.62^{+0.03}_{-0.03}$ & $~\,~~0.86^{+0.10}_{-0.10}$ \\ 
NGC1052 & $40.36^{+0.02}_{-0.01}$ & $41.95^{+0.02}_{-0.01}$ & $41.53^{+0.03}_{-0.03}$ & $41.87^{+0.01}_{-0.01}$ & $~\,~~42.71^{+0.04}_{-0.04}$ & $~~\,~42.79^{+0.03}_{-0.03}$ & $~\,~~0.83^{+0.10}_{-0.10}$ \\ 
2MFGC 2280 & $39.54^{+0.11}_{-0.06}$ & $42.56^{+0.07}_{-0.02}$ & $42.79^{+0.16}_{-0.15}$ & $43.10^{+0.11}_{-0.11}$ & $<43.20$ & $~~\,~43.74^{+0.02}_{-0.02}$ & $<0.19$ \\ 
NGC1365 & $40.56^{+0.02}_{-0.01}$ & $42.43^{+0.01}_{-0.02}$ & $41.77^{+0.04}_{-0.04}$ & $41.84^{+0.03}_{-0.03}$ & $<44.09$ & $~~\,~44.60^{+0.02}_{-0.02}$ & $<0.26$ \\ 
2MASXJ04234080+0408017 & $40.81^{+0.13}_{-0.01}$ & $43.84^{+0.02}_{-0.02}$ & $43.83^{+0.08}_{-0.09}$ & $43.99^{+0.06}_{-0.06}$ & $~\,~~44.30^{+0.09}_{-0.13}$ & $~~\,~44.50^{+0.04}_{-0.03}$ & $~\,~~0.64^{+0.14}_{-0.11}$ \\ 
CGCG420-015 & $40.73^{+0.02}_{-0.02}$ & $43.53^{+0.02}_{-0.01}$ & $43.46^{+0.09}_{-0.09}$ & $43.54^{+0.08}_{-0.08}$ & $~\,~~44.32^{+0.06}_{-0.07}$ & $~~\,~44.42^{+0.04}_{-0.05}$ & $~\,~~0.79^{+0.10}_{-0.10}$ \\ 
ESO 033-G002 & $41.37^{+0.01}_{-0.02}$ & $43.15^{+0.01}_{-0.02}$ & $42.95^{+0.07}_{-0.07}$ & $42.83^{+0.05}_{-0.05}$ & $~\,~~43.82^{+0.06}_{-0.06}$ & $~~\,~43.93^{+0.03}_{-0.03}$ & $~\,~~0.77^{+0.10}_{-0.11}$ \\ 
LEDA178130 & $41.66^{+0.03}_{-0.01}$ & $44.02^{+0.07}_{-0.01}$ & $43.82^{+0.03}_{-0.03}$ & $44.05^{+0.01}_{-0.03}$ & $~\,~~43.91^{+0.06}_{-0.05}$ & $~~\,~43.97^{+0.04}_{-0.04}$ & $~\,~~0.86^{+0.10}_{-0.10}$ \\ 
2MASXJ05081967+1721483 & $41.27^{+0.01}_{-0.01}$ & $43.16^{+0.02}_{-0.02}$ & $42.86^{+0.06}_{-0.06}$ & $43.05^{+0.04}_{-0.04}$ & $~\,~~43.47^{+0.23}_{-0.46}$ & $~~\,~44.08^{+0.03}_{-0.03}$ & $~\,~~0.24^{+0.22}_{-0.15}$ \\ 
NGC2110 & $42.24^{+0.02}_{-0.01}$ & $43.79^{+0.01}_{-0.01}$ & $43.60^{+0.01}_{-0.02}$ & $43.85^{+0.01}_{-0.02}$ & $~\,~~43.12^{+0.12}_{-0.21}$ & $~~\,~43.80^{+0.02}_{-0.03}$ & $~\,~~0.20^{+0.10}_{-0.10}$ \\ 
ESO 005-G004 & $39.51^{+0.01}_{-0.07}$ & $41.74^{+0.02}_{-0.05}$ & $41.41^{+0.51}_{-0.44}$ & $41.55^{+0.47}_{-0.39}$ & $<42.68$ & $~~\,~43.68^{+0.01}_{-0.01}$ & $<0.10$ \\ 
ESO 121-IG028 & $41.14^{+0.01}_{-0.02}$ & $43.72^{+0.02}_{-0.01}$ & $43.53^{+0.08}_{-0.08}$ & $43.65^{+0.05}_{-0.05}$ & $~\,~~43.36^{+0.17}_{-0.17}$ & $~~\,~43.63^{+0.06}_{-0.07}$ & $~\,~~0.53^{+0.11}_{-0.13}$ \\ 
MCG+06-16-028 & $39.98^{+0.07}_{-0.05}$ & $42.66^{+0.05}_{-0.03}$ & $42.86^{+0.16}_{-0.15}$ & $43.01^{+0.12}_{-0.13}$ & $~\,~~43.56^{+0.10}_{-0.16}$ & $~~\,~44.09^{+0.02}_{-0.03}$ & $~\,~~0.30^{+0.10}_{-0.10}$ \\ 
LEDA96373 & $40.68^{+0.02}_{-0.02}$ & $43.38^{+0.02}_{-0.02}$ & $43.20^{+0.18}_{-0.22}$ & $43.23^{+0.17}_{-0.21}$ & $>44.57$ & $<44.61$ & $>0.90$ \\ 
UGC3995A & $40.54^{+0.02}_{-0.01}$ & $42.99^{+0.02}_{-0.02}$ & $42.55^{+0.09}_{-0.10}$ & $42.74^{+0.07}_{-0.08}$ & $~\,~~43.34^{+0.24}_{-0.28}$ & $~~\,~43.85^{+0.05}_{-0.05}$ & $~\,~~0.30^{+0.10}_{-0.10}$ \\ 
Mrk 1210 & $41.04^{+0.02}_{-0.02}$ & $43.20^{+0.02}_{-0.02}$ & $42.90^{+0.05}_{-0.05}$ & $42.99^{+0.04}_{-0.04}$ & $~\,~~44.08^{+0.08}_{-0.09}$ & $~~\,~44.15^{+0.06}_{-0.07}$ & $~\,~~0.84^{+0.10}_{-0.10}$ \\ 
MCG-01-22-006 & $41.10^{+0.01}_{-0.01}$ & $43.49^{+0.02}_{-0.01}$ & $43.14^{+0.06}_{-0.06}$ & $43.45^{+0.04}_{-0.04}$ & $~\,~~42.98^{+0.21}_{-0.26}$ & $~~\,~43.87^{+0.02}_{-0.02}$ & $~\,~~0.13^{+0.10}_{-0.10}$ \\ 
MCG+11-11-032 & $41.39^{+0.04}_{-0.02}$ & $43.83^{+0.04}_{-0.02}$ & $43.53^{+0.15}_{-0.17}$ & $43.63^{+0.10}_{-0.13}$ & $~\,~~43.50^{+0.07}_{-0.07}$ & $~~\,~43.78^{+0.02}_{-0.03}$ & $~\,~~0.52^{+0.10}_{-0.10}$ \\ 
Mrk 18 & $40.12^{+0.68}_{-0.02}$ & $42.02^{+0.92}_{-0.01}$ & $41.76^{+0.15}_{-0.13}$ & $42.02^{+0.06}_{-0.10}$ & $~\,~~43.40^{+0.24}_{-0.35}$ & $~~\,~43.68^{+0.03}_{-0.03}$ & $~\,~~0.52^{+0.29}_{-0.44}$ \\ 
IC 2461 & $40.62^{+0.01}_{-0.02}$ & $42.02^{+0.03}_{-0.01}$ & $42.23^{+0.09}_{-0.09}$ & $42.36^{+0.06}_{-0.06}$ & $~\,~~42.42^{+0.25}_{-0.21}$ & $~~\,~43.03^{+0.01}_{-0.02}$ & $~\,~~0.25^{+0.10}_{-0.15}$ \\ 
MCG-01-24-012 & $41.52^{+0.02}_{-0.01}$ & $43.51^{+0.01}_{-0.01}$ & $43.38^{+0.06}_{-0.06}$ & $43.33^{+0.04}_{-0.04}$ & $~\,~~43.75^{+0.07}_{-0.07}$ & $~~\,~43.93^{+0.04}_{-0.04}$ & $~\,~~0.66^{+0.10}_{-0.10}$ \\ 
2MASXJ09235371-3141305 & $40.80^{+0.03}_{-0.03}$ & $43.77^{+0.04}_{-0.03}$ & $43.72^{+0.14}_{-0.15}$ & $43.82^{+0.08}_{-0.09}$ & $~\,~~42.92^{+0.20}_{-0.34}$ & $~~\,~43.53^{+0.03}_{-0.03}$ & $~\,~~0.24^{+0.17}_{-0.11}$ \\ 
NGC2992 & $41.67^{+0.02}_{-0.01}$ & $43.14^{+0.01}_{-0.01}$ & $42.87^{+0.02}_{-0.02}$ & $43.07^{+0.02}_{-0.01}$ & $<43.57$ & $~~\,~43.91^{+0.02}_{-0.02}$ & $<0.40$ \\ 
NGC3079 & $39.45^{+0.02}_{-0.09}$ & $41.72^{+0.01}_{-0.03}$ & $43.09^{+0.16}_{-0.14}$ & $43.08^{+0.14}_{-0.12}$ & $<43.55$ & $~~\,~44.55^{+0.01}_{-0.01}$ & $<0.10$ \\ 
ESO 263-G013 & $40.67^{+0.08}_{-0.02}$ & $43.52^{+0.08}_{-0.02}$ & $43.57^{+0.07}_{-0.08}$ & $43.76^{+0.04}_{-0.06}$ & $>43.82$ & $<44.00$ & $>0.80$ \\ 
NGC3281 & $40.35^{+0.01}_{-0.02}$ & $42.86^{+0.02}_{-0.02}$ & $42.22^{+0.22}_{-0.26}$ & $42.49^{+0.22}_{-0.26}$ & $~\,~~43.83^{+0.11}_{-0.14}$ & $~~\,~44.26^{+0.03}_{-0.04}$ & $~\,~~0.38^{+0.10}_{-0.10}$ \\ 
MCG+12-10-067 & $40.77^{+0.03}_{-0.02}$ & $43.20^{+0.03}_{-0.02}$ & $42.95^{+0.14}_{-0.15}$ & $43.01^{+0.09}_{-0.11}$ & $~\,~~44.02^{+0.10}_{-0.11}$ & $~~\,~44.43^{+0.02}_{-0.02}$ & $~\,~~0.39^{+0.10}_{-0.10}$ \\ 
MCG+06-24-008 & $40.83^{+0.02}_{-0.01}$ & $42.98^{+0.01}_{-0.02}$ & $42.67^{+0.04}_{-0.04}$ & $42.98^{+0.02}_{-0.02}$ & $<42.98$ & $~~\,~43.98^{+0.01}_{-0.02}$ & $<0.10$ \\ 
UGC5881 & $40.49^{+0.03}_{-0.01}$ & $42.72^{+0.05}_{-0.01}$ & $42.33^{+0.15}_{-0.15}$ & $42.59^{+0.09}_{-0.11}$ & $~\,~~43.66^{+0.18}_{-0.24}$ & $~~\,~44.13^{+0.02}_{-0.02}$ & $~\,~~0.34^{+0.17}_{-0.17}$ \\ 
NGC3393 & $39.85^{+0.06}_{-0.08}$ & $42.76^{+0.04}_{-0.03}$ & $43.50^{+0.15}_{-0.13}$ & $43.61^{+0.11}_{-0.09}$ & $~\,~~43.64^{+0.07}_{-0.07}$ & $~~\,~43.93^{+0.03}_{-0.03}$ & $~\,~~0.51^{+0.10}_{-0.10}$ \\ 
Mrk 728 & $41.16^{+0.02}_{-0.02}$ & $43.34^{+0.02}_{-0.02}$ & $43.04^{+0.04}_{-0.04}$ & $43.33^{+0.02}_{-0.01}$ & $>42.91$ & $<43.28$ & $>0.49$ \\ 
2MASXJ11364205-6003070 & $40.91^{+0.01}_{-0.01}$ & $42.60^{+0.01}_{-0.02}$ & $42.36^{+0.08}_{-0.09}$ & $42.36^{+0.06}_{-0.07}$ & $~\,~~42.91^{+0.24}_{-0.38}$ & $~~\,~43.63^{+0.03}_{-0.03}$ & $~\,~~0.20^{+0.14}_{-0.11}$ \\ 
NGC3786 & $40.09^{+0.05}_{-0.02}$ & $41.65^{+0.07}_{-0.03}$ & $41.49^{+0.18}_{-0.18}$ & $41.66^{+0.11}_{-0.13}$ & $<42.76$ & $~~\,~43.51^{+0.03}_{-0.03}$ & $<0.16$ \\ 
NGC4388 & $40.63^{+0.01}_{-0.02}$ & $42.82^{+0.01}_{-0.01}$ & $42.27^{+0.02}_{-0.02}$ & $42.48^{+0.02}_{-0.02}$ & $~\,~~43.00^{+0.16}_{-0.18}$ & $~~\,~43.58^{+0.03}_{-0.03}$ & $~\,~~0.26^{+0.10}_{-0.11}$ \\ 
LEDA170194 & $41.00^{+0.03}_{-0.01}$ & $43.26^{+0.09}_{-0.02}$ & $43.02^{+0.07}_{-0.07}$ & $43.22^{+0.04}_{-0.05}$ & $~\,~~43.85^{+0.09}_{-0.11}$ & $~~\,~44.16^{+0.02}_{-0.03}$ & $~\,~~0.48^{+0.10}_{-0.13}$ \\ 
NGC4941 & $39.40^{+0.18}_{-0.03}$ & $41.39^{+0.13}_{-0.02}$ & $41.73^{+0.16}_{-0.16}$ & $41.91^{+0.11}_{-0.12}$ & $~\,~~42.36^{+0.05}_{-0.05}$ & $~~\,~42.69^{+0.02}_{-0.02}$ & $~\,~~0.47^{+0.10}_{-0.10}$ \\ 
NGC4992 & $40.86^{+0.01}_{-0.02}$ & $43.55^{+0.02}_{-0.01}$ & $43.22^{+0.05}_{-0.06}$ & $43.52^{+0.04}_{-0.05}$ & $~\,~~43.55^{+0.07}_{-0.06}$ & $~~\,~43.77^{+0.03}_{-0.03}$ & $~\,~~0.59^{+0.10}_{-0.10}$ \\ 
Mrk 248 & $41.41^{+0.02}_{-0.01}$ & $43.74^{+0.02}_{-0.01}$ & $43.55^{+0.10}_{-0.10}$ & $43.56^{+0.07}_{-0.07}$ & $~\,~~44.21^{+0.14}_{-0.40}$ & $~~\,~44.59^{+0.03}_{-0.04}$ & $~\,~~0.41^{+0.29}_{-0.14}$ \\ 
ESO 509-IG066 & $41.43^{+0.02}_{-0.01}$ & $43.77^{+0.04}_{-0.02}$ & $43.47^{+0.11}_{-0.11}$ & $43.68^{+0.07}_{-0.07}$ & $~\,~~43.97^{+0.15}_{-0.19}$ & $~~\,~44.46^{+0.04}_{-0.03}$ & $~\,~~0.34^{+0.12}_{-0.11}$ \\ 
NGC5252 & $41.46^{+0.02}_{-0.01}$ & $43.49^{+0.02}_{-0.02}$ & $43.25^{+0.03}_{-0.02}$ & $43.49^{+0.02}_{-0.01}$ & $~\,~~43.54^{+0.07}_{-0.11}$ & $~~\,~43.80^{+0.03}_{-0.03}$ & $~\,~~0.54^{+0.11}_{-0.10}$ \\ 
NGC5273 & $40.82^{+0.01}_{-0.02}$ & $42.05^{+0.02}_{-0.02}$ & $41.74^{+0.05}_{-0.04}$ & $41.88^{+0.03}_{-0.03}$ & $~\,~~41.48^{+0.19}_{-0.27}$ & $~~\,~42.25^{+0.02}_{-0.02}$ & $~\,~~0.15^{+0.11}_{-0.10}$ \\ 
NGC5674 & $41.34^{+0.01}_{-0.02}$ & $43.35^{+0.03}_{-0.01}$ & $43.18^{+0.08}_{-0.08}$ & $43.27^{+0.05}_{-0.06}$ & $<43.38$ & $~~\,~44.38^{+0.01}_{-0.01}$ & $<0.10$ \\ 
NGC5728 & $40.10^{+0.02}_{-0.01}$ & $42.93^{+0.01}_{-0.02}$ & $42.84^{+0.04}_{-0.04}$ & $43.10^{+0.03}_{-0.03}$ & $~\,~~42.97^{+0.27}_{-0.37}$ & $~~\,~43.74^{+0.02}_{-0.02}$ & $~\,~~0.18^{+0.14}_{-0.14}$ \\ 
IC 4518A & $40.87^{+0.03}_{-0.01}$ & $43.04^{+0.02}_{-0.02}$ & $42.68^{+0.10}_{-0.11}$ & $42.68^{+0.08}_{-0.09}$ & $~\,~~44.03^{+0.17}_{-0.30}$ & $~~\,~44.41^{+0.03}_{-0.04}$ & $~\,~~0.41^{+0.23}_{-0.23}$ \\ 
2MASXJ15064412+0351444 & $40.83^{+0.02}_{-0.02}$ & $43.01^{+0.03}_{-0.03}$ & $42.80^{+0.05}_{-0.05}$ & $43.02^{+0.03}_{-0.03}$ & $<42.97$ & $~~\,~43.26^{+0.03}_{-0.03}$ & $<0.42$ \\ 
NGC5899 & $40.71^{+0.02}_{-0.01}$ & $42.41^{+0.01}_{-0.02}$ & $42.20^{+0.08}_{-0.07}$ & $42.27^{+0.05}_{-0.05}$ & $<43.46$ & $~~\,~44.09^{+0.02}_{-0.01}$ & $<0.19$ \\ 
MCG+11-19-006 & $40.81^{+0.06}_{-0.01}$ & $43.49^{+0.07}_{-0.02}$ & $43.13^{+0.13}_{-0.14}$ & $43.43^{+0.08}_{-0.10}$ & $~\,~~43.84^{+0.11}_{-0.13}$ & $~~\,~44.31^{+0.03}_{-0.03}$ & $~\,~~0.35^{+0.10}_{-0.10}$ \\ 
MCG-01-40-001 & $41.02^{+0.01}_{-0.02}$ & $43.14^{+0.02}_{-0.01}$ & $42.82^{+0.08}_{-0.08}$ & $42.97^{+0.05}_{-0.06}$ & $~\,~~44.06^{+0.05}_{-0.07}$ & $~~\,~44.31^{+0.02}_{-0.02}$ & $~\,~~0.55^{+0.10}_{-0.10}$ \\ 
NGC5995 & $41.60^{+0.02}_{-0.01}$ & $43.53^{+0.01}_{-0.02}$ & $43.32^{+0.04}_{-0.04}$ & $43.33^{+0.03}_{-0.03}$ & $~\,~~44.29^{+0.11}_{-0.13}$ & $~~\,~44.76^{+0.02}_{-0.02}$ & $~\,~~0.34^{+0.10}_{-0.10}$ \\ 
MCG+14-08-004 & $40.61^{+0.06}_{-0.02}$ & $42.83^{+0.23}_{-0.02}$ & $42.64^{+0.10}_{-0.10}$ & $42.85^{+0.04}_{-0.08}$ & $~\,~~43.12^{+0.08}_{-0.09}$ & $~~\,~43.28^{+0.04}_{-0.05}$ & $~\,~~0.68^{+0.10}_{-0.11}$ \\ 
NGC6240 & $40.79^{+0.04}_{-0.05}$ & $43.69^{+0.01}_{-0.02}$ & $43.81^{+0.05}_{-0.04}$ & $44.02^{+0.04}_{-0.02}$ & $<45.21$ & $~~\,~45.36^{+0.02}_{-0.03}$ & $<0.67$ \\ 
NGC6300 & $40.79^{+0.02}_{-0.01}$ & $42.30^{+0.02}_{-0.01}$ & $41.99^{+0.03}_{-0.03}$ & $42.06^{+0.02}_{-0.02}$ & $~\,~~43.09^{+0.13}_{-0.15}$ & $~~\,~43.65^{+0.02}_{-0.02}$ & $~\,~~0.27^{+0.10}_{-0.10}$ \\ 
MCG+07-37-031 & $41.65^{+0.01}_{-0.02}$ & $43.99^{+0.03}_{-0.02}$ & $43.69^{+0.06}_{-0.07}$ & $43.91^{+0.04}_{-0.05}$ & $~\,~~43.78^{+0.18}_{-0.28}$ & $~~\,~44.18^{+0.03}_{-0.03}$ & $~\,~~0.39^{+0.21}_{-0.20}$ \\ 
IC 4709 & $40.81^{+0.02}_{-0.01}$ & $43.08^{+0.02}_{-0.01}$ & $42.78^{+0.08}_{-0.08}$ & $42.83^{+0.06}_{-0.06}$ & $~\,~~43.49^{+0.06}_{-0.06}$ & $~~\,~43.71^{+0.03}_{-0.03}$ & $~\,~~0.61^{+0.10}_{-0.10}$ \\ 
ESO 103-G035 & $41.43^{+0.01}_{-0.02}$ & $43.46^{+0.01}_{-0.02}$ & $43.28^{+0.02}_{-0.02}$ & $43.31^{+0.02}_{-0.01}$ & $~\,~~44.09^{+0.08}_{-0.10}$ & $~~\,~44.18^{+0.06}_{-0.07}$ & $~\,~~0.81^{+0.10}_{-0.10}$ \\ 
2MASXJ20183871+4041003 & $40.80^{+0.02}_{-0.01}$ & $42.94^{+0.02}_{-0.01}$ & $42.58^{+0.08}_{-0.09}$ & $42.79^{+0.05}_{-0.06}$ & $>43.12$ & $<43.37$ & $>0.75$ \\ 
MCG+04-48-002 & $40.06^{+0.10}_{-0.01}$ & $42.60^{+0.03}_{-0.01}$ & $42.56^{+0.14}_{-0.15}$ & $42.73^{+0.09}_{-0.11}$ & $<43.42$ & $~~\,~44.42^{+0.02}_{-0.02}$ & $<0.10$ \\ 
IC 5063 & $41.02^{+0.02}_{-0.02}$ & $43.10^{+0.02}_{-0.02}$ & $42.87^{+0.05}_{-0.05}$ & $43.01^{+0.03}_{-0.03}$ & $~\,~~44.26^{+0.06}_{-0.05}$ & $~~\,~44.33^{+0.05}_{-0.04}$ & $~\,~~0.84^{+0.10}_{-0.10}$ \\ 
MCG+06-49-019 & $40.37^{+0.03}_{-0.02}$ & $42.26^{+0.07}_{-0.03}$ & $41.97^{+0.21}_{-0.21}$ & $42.18^{+0.03}_{-0.03}$ & $~\,~~42.95^{+0.11}_{-0.11}$ & $~~\,~43.40^{+0.02}_{-0.02}$ & $~\,~~0.35^{+0.10}_{-0.10}$ \\ 
MCG+01-57-016 & $41.04^{+4.41}_{-0.02}$ & $43.06^{+0.02}_{-0.01}$ & $42.73^{+0.05}_{-0.05}$ & $42.84^{+0.04}_{-0.04}$ & $~\,~~43.89^{+0.11}_{-0.12}$ & $~~\,~44.14^{+0.03}_{-0.03}$ & $~\,~~0.57^{+0.12}_{-0.16}$ \\ 
NGC7582 & $40.41^{+0.02}_{-0.02}$ & $42.40^{+0.02}_{-0.02}$ & $41.61^{+0.07}_{-0.07}$ & $41.77^{+0.07}_{-0.07}$ & $<43.84$ & $~~\,~44.29^{+0.02}_{-0.02}$ & $<0.29$ \\ 
2MASXJ23303771+7122464 & $40.83^{+0.33}_{-0.01}$ & $43.20^{+0.13}_{-0.02}$ & $42.90^{+0.16}_{-0.15}$ & $43.14^{+0.08}_{-0.11}$ & $<43.30$ & $~~\,~44.04^{+0.02}_{-0.02}$ & $<0.16$ \\ 
PKS 2331-240 & $41.80^{+0.01}_{-0.02}$ & $43.93^{+0.01}_{-0.01}$ & $43.79^{+0.02}_{-0.02}$ & $43.92^{+0.01}_{-0.01}$ & $~\,~~43.98^{+0.04}_{-0.03}$ & $~~\,~44.12^{+0.03}_{-0.02}$ & $~\,~~0.71^{+0.10}_{-0.10}$ 
\enddata
\tablecomments{Observed and intrinsic X-ray luminosities in the 2--10\,keV and 10--50\,keV bands derived from the X-ray fitting (Columns 2-5). Intrinsic luminosities are corrected for the effects of reflection as well as absorption.  Reflected luminosities are calculated by multiplying the reflection parameter by the intrinsic luminosity. The AGN component of the IR luminosity from the decomposition of the SED is given in Column 6. Its fraction relative to the total IR luminosity from the fits (Column 7) is given in Column 8.}
\end{deluxetable*}

\end{document}